\documentclass[showpacs,preprint]{revtex4}

\usepackage{graphics}
\usepackage{graphicx}

\def\m@thcombine#1#2{
 \setbox0=\hbox{$#1$}
 \setbox1=\hbox{$#2$}
 \ifdim\wd0>\wd1
  \setbox0=\hbox to\wd1{\hss\box0\hss}
 \else
  \setbox1=\hbox to\wd0{\hss\box1\hss}
 \fi \mathop{\vcenter{
  \offinterlineskip\box0\box1}}}
\def\lesim{\m@thcombine<\sim}
\def\gesim{\m@thcombine>\sim}

\newcommand{\vecr}{\mbox{\boldmath $r$}}
\newcommand{\vecrp}{\mbox{\boldmath $r$}'}

\newcommand{\eps}{\epsilon}

\newcommand{\rp}{\mbox{$r$}'}

\newcommand{\gtsim}{\protect\raisebox{-0.8ex}{$\:\stackrel{\textstyle >}{\sim}\:$}}

\begin{document}

\title{
 Di-neutron correlation in monopole two-neutron transfer modes
 in Sn isotope chain
}

\author{Hirotaka Shimoyama, Masayuki Matsuo}

\affiliation{
 Department of Physics, Faculty of Science and 
 Graduate School of Science and Technology, 
 Niigata University, Niigata 950-2181, Japan}

\date{\today}

\begin{abstract}
 We study microscopic structures of monopole pair vibrational
 modes and associated two-neutron transfer amplitudes 
 in neutron-rich Sn isotopes 
 by means of the linear response formalism of the
 quasiparticle random phase approximation(QRPA).
 For this purpose we introduce a method to decompose the transfer
 amplitudes with respect to two-quasiparticle components of the QRPA eigen
 mode.
 It is found that 
 pair-addition vibrational modes 
 in neutron-rich $^{132-140}$Sn
 and the pair rotational modes 
 in $^{142-150}$Sn are commonly characterized by 
 coherent contributions of quasaiparticle states having
 high orbital angular momenta $l \gesim 5$, which suggests
 transfer of a spatially correlated neutron pair.
 The calculation also predicts a high-lying pair vibration,
 the giant pair vibration, emerging near the one-neutron separation
 energy in $^{110-130}$Sn, and we find that they
 have the same di-neutron characters as that of 
 the low-lying pair vibration in $^{132-140}$Sn.
\end{abstract}


\maketitle

\section{Introduction \label{Intro}}
 Nuclei close to the drip-lines often exhibit exotic features
 which are not present in stable nuclei.
 Representative examples are the neutron halo and 
 skin\cite{Tanihata1985,TanihataReview}, which are associated with 
 the weak binding of the last neutrons.
 Another example is the neutron pair correlation,
 which may be enhanced in the surface region
 with the skin or halo\cite{Bertsch91,Dobaczewski01,MatsuoJPG10}. The
 neutron pairing in the low density surface region
 may be related to the predicted strong density dependence of the
 pairing in neutron
 matter\cite{Lombardo01,Dean03,Matsuo06,Margueron08,Gezerlis10}.

 Because of the low-density enhancement,
 the pair correlation in neutron-rich nuclei may lead to 
 the di-neutron correlation,
 i.e., a strong spatial correlation among the two neutrons of 
 the neutron Cooper pair. The di-neutron correlation has been discussed 
 in the two-neutron halo nuclei, e.g.
 $^{11}$Li\cite{Bertsch91,Hansen87,Esbessen92,Ikeda92,Zhukov93,Hagino05,Aoyama06,En'yo07,Myo08}, and
 recently its possibility is extended for heavier-mass pair correlated
 nuclei\cite{Matsuo05,Pillet07,Pillet10,Matsuo12PhysScr,MatsuoBCS}.
 The spatial correlation has been discussed also
 for the closed-shell plus two-neutron (two-neutron hole) configuration,
 e.g. $^{210}$Pb\cite{Ibarra77,Catara84,Ferreira84}.
 It is argued also that the di-neutron correlation affects the nature of
 excitation modes, a typical example of which is the soft dipole
 excitation in the two-neutron halo nucleus
 $^{11}$Li\cite{Bertsch91,Hagino05,Myo08,Nakamura06}.
 It is also predicted that the soft octupole excitation in medium heavy
 neutron-rich nuclei, e.g. $^{84}$Ni\cite{Serizawa09},
 and excited states of clustering nuclei,
 e.g. $^{10}$Be\cite{Kobayashi12}, exhibit the di-neutron feature.

 The most direct influence of the pair correlation is expected to be
 seen in transfers of two nucleons. 
 The pair correlation induces 
 collective excitation modes, known as the pair
 vibration and the pair rotation, which accompany
 enhanced two-nucleon transfer reaction cross
 sections\cite{Bes-Broglia66,Broglia73,Brink-Broglia,Yoshida1962,BohrVorII}.
 Recently, the two-neutron transfer experiments 
 have been conducted for some neutron-rich
 nuclei\cite{Keeley07,Tanihata08,Chatterjee08,Golovkov09,Wimmer11,Lemasson11}. 
 Motivated by the above mentioned
 interests and prospects of future experiments,
 theoretical studies of the two-neutron transfers 
 in neutron rich nuclei also have been
 performed\cite{Khan04,Avez08,Khan09,Matsuo10,Shimoyama11,Potel11,Grasso12,Potel13}.

 We have been investigating the two-neutron transfer modes
 in neutron-rich Sn isotopes in order to explore possible new features
 of the pair vibration and the pair
 rotation\cite{Matsuo10,Shimoyama11}. 
 We adopt the 
 nuclear energy density functional model, in particular, 
 the Skyrme-Hartree-Fock-Bogoliubov model to describe the
 pair correlated ground state and the associated pair rotational
 two-neutron transfer mode, and we apply the quasiparticle
 random phase approximation (QRPA) to study the pair vibrations
 and the associated two-neutron transfer amplitudes. In a 
 preceding work\cite{Shimoyama11}, where we described 
 monopole low-lying pair vibration in the whole isotope chain
 $100 \le A \le 150$, we found 
 a characteristic pair vibration in $132 \le A \le 140$, which
 has a large pair-addition strength, comparable
 to that of the ground-state pair-rotational mode, 
 and much larger than the independent-particle transitions.  
 An interesting feature is that the
 transition density of this pair vibration mode has a long
 tail which reaches far outside the nuclear
 surface $r \sim 15$ fm.

 One of the purposes of the present paper is to clarify
 the nature of the characteristic pair vibration predicted in our
 previous study\cite{Shimoyama11} for neutron-rich Sn $A=132-140$ isotopes.
 We shall analyze in detail the microscopic structure of
 this mode by looking into the single-particle compositions. 
 We decompose the 
 transition density and the QRPA phonon amplitude 
 with respect to two-quasiparticle configurations forming 
 the QRPA phonon. As shown later, we will find that a large number of 
 both weakly bound and unbound quasi-neutron states contribute 
 coherently to build up the characteristic correlation in this mode.
 We will find further that this correlation is consistent with
 what we expect for a spatially correlated neutron-pair, the
 di-neutron, which is transfered in the pair vibration. 

 In the previous paper\cite{Shimoyama11}, we found also that the pair-transfer
 strength associated with the pair rotation (the transition between
 the ground states of the neighboring
 even nuclei) significantly increases as the neutron
 number further increases, i.e., in the isotopes beyond
 $A= 140$. In the present paper, we shall also analyze microscopic
 structure of this enhanced ground state transfer. We will show that
 it is indeed related to the characteristic pair vibration
 in $^{132-140}$Sn. In the present study, we also
 aim at widening the scope of our study of the pairing collectivity.
 Namely, we perform our numerical calculation in a very wide 
 interval $A= 100\sim150$ of the Sn isotopic chain, 
 including the stable isotopes, and we explore
 high excitation energy region $\sim 10-20$ MeV, where
 we can expect the giant pair vibration. 
 We will discuss how the giant pair vibration, which we find in the
 isotopes $A=110-130$, is related to the low-lying pair vibration and the
 pair rotation in the very neutron-rich nuclei $A= 132-140$ 
 and $A> 140$.

\section{Two-quasiparticle components of the QRPA mode}

\subsection{Linear response formulation of the QRPA}

 In the present study, we adopt the Skyrme-Hartree-Fock-Bogoliubov model 
 to describe the pair correlated ground states, and apply
 a linear response formulation of
 the quasiparticle random phase approximation (QRPA) 
 to describe 
 the excitation modes\cite{Matsuo01,Serizawa09,Matsuo10,Shimoyama11}.
 The framework is essentially the same as what we have
 adopted in the preceding study of the two-neutron transfer 
 in neutron-rich Sn isotopes\cite{Matsuo10,Shimoyama11}.
 Here we recapitulate only the basic framework with emphasis
 on new aspects of the formulation in the present work.

 We describe the nuclear response at the frequency $\omega$ by
 solving the linear response equation
\begin{eqnarray}
\delta \rho_{\alpha L}(r,\omega)~=~
\int_0 d \rp \sum_{\beta}R_{0, L}^{\alpha\beta}(r,\rp, \omega)
\left(
\sum_{\gamma}\frac{\partial v_{\beta}}
{\partial \rho_{\gamma}}(\rp )
{1 \over \rp ^{2}} \delta\rho_{\gamma L}(\rp, \omega)
+v^{ext}_{\beta L}(\rp)
\right)
\label{eq:drho}
\end{eqnarray}
 for three kinds of densities 
$\delta \rho_{\alpha L}(r,\omega)$=$\delta \rho_{L}(r,\omega)$ and 
$\delta {\tilde \rho}_{\pm L}(r,\omega)$, which are fluctuating parts 
of the normal density and the two kinds of pair densities;
\begin{eqnarray}
 \rho(\vecr)&=&
 \langle 0|
{\hat \rho}(\vecr)
|0 \rangle \label{eq:normalD}\\
&=& \langle 0|
\sum_{\sigma}\psi^{\dagger}(\vecr\sigma)\psi(\vecr\sigma)
|0 \rangle,\\
{\tilde \rho}_{\pm}(\vecr)&=&
 \langle 0|
 \hat{{\tilde \rho}}_{\pm}(\vecr)
|0 \rangle \label{eq:pairDs}\\
&=& \langle 0|
\psi^{\dagger}(\vecr \downarrow)\psi^{\dagger}(\vecr\uparrow)
\pm
\psi^{~}(\vecr\uparrow)\psi^{~}(\vecr\downarrow)
|0 \rangle.
\end{eqnarray}

 The unperturbed response function $R_{0,q L}^{\alpha\beta}(r,r',\omega)$
 is expressed as
\begin{eqnarray}
&~&R_{0, L}^{\alpha\beta}(r,r',\omega)~=~\frac{1}{2}\sum_{i\geq i'}
\frac{|\langle l'j'|| Y_{ L} || lj\rangle| ^{2}}
{2L+1}\nonumber\\
&~&~~~~~~~~~~~~~~~~~~~~~~~\times
\left\{
\frac{
\langle 0|\rho_{\alpha}(r)| ii'\rangle
\langle ii'|\rho_{\beta}(r')| 0\rangle}
{\hbar\omega +i\eps -E_{i}-E_{i'}}
-\frac{
\langle 0|\rho_{\beta}(r')| ii'\rangle
\langle ii'|\rho_{\alpha}(r)| 0\rangle
}{\hbar\omega +i\eps +E_{i}+E_{i'}}
\right\},
\label{eq:spR0F}
\end{eqnarray}
 with
\begin{eqnarray}
\langle ii'|\rho_{\alpha}(r)|0 \rangle
&\equiv&
\phi^{\rm T}_{i}(r){\cal A}{\bar \phi}_{i'}(r), \label{eq:ii0}
\\
\langle 0|\rho_{\alpha}(r)|ii' \rangle
&\equiv&
 {\bar \phi}^{\rm T}_{i'}(r){\cal A} \phi_{i}(r), \label{eq:0ii}
\end{eqnarray}
 if we adopt the spectral representation.
 Here ${\cal A}$ is $2\times 2$ matrices which correspond
 to the three kinds of densities ${\rho}_{\alpha}(\vecr)$= 
 ${ \rho}(\vecr)$ and ${ {\tilde\rho}}_{\pm }(\vecr)$. 
 Index $i$ is a quantum number $[nlj]_q$ ($q$=n, p) of the
 one-quasiparticle state whose energy is $E_i$ and wave function is
 $\phi_i(\vecr\sigma)=Y_{ljm}(\hat{\vecr}\sigma)\phi_i(r)/r$ where
 $\phi_{i}(\hat{\vecr}\sigma)$ and its radial part $\phi_{i}(r)$ have 
 two components, 
 $\phi_{i}(r)=\left[\varphi_{1,i}(r),\varphi_{2,i}(r)\right]^{T}$.

 In the preceding paper\cite{Shimoyama11}, we adopted 
 the continuum version of the linear response QRPA, i.e. the
 continuum QRPA\cite{Matsuo01}, in which the unperturbed response
 function is constructed in terms of the HFB Green's function in place
 of the spectral representation. The continuum QRPA is appropriate to
 describe the response at energies above the particle-emission
 threshold.
 However in this present work, we  use the unperturbed response function
 of Eq. (\ref{eq:spR0F})
 together with the discretized quasiparticle eigenstates of
 the HFB equation.
 As shown below, this enables us to express the microscopic structure
 of the QRPA phonon and the transition densities in terms of
 two-quasiparticle components.

\subsection{Two-quasiparticle components of the QRPA phonon}

 It is customary to expresses the QRPA phonon operator as a
 superposition of two-quasiparticle creation and annihilation operators:
\begin{eqnarray}
O_{LM}^{\nu \dagger}&=&
\sum_{i\geq i'}{1 \over \sqrt{1+\delta_{ii'}}}
\left\{
 X^{\nu}_{ii'}
\left[a_{i}^{\dagger}a_{i'}^{\dagger}\right]_{LM}
-Y^{\nu}_{ii'}
\left[a_{i'}^{~}a_{i}^{~}\right]_{LM}
\right\},
\end{eqnarray}
 with the two-quasiparticle operators defined by
\begin{eqnarray}
\left[a_{i}^{\dagger}a_{i'}^{\dagger}\right]_{LM} \equiv
\sum_{mm'}\langle
jmj'm'|LM\rangle a_{nljm}^{\dagger}a_{n'l'j'm'}^{\dagger},
\\
\left[a_{i}^{~}a_{i'}^{~}\right]_{LM} \equiv \sum_{mm'}
\langle jmj'm'|LM\rangle
{\tilde a}_{nljm}{\tilde
 a}_{n'l'j'm'},
\end{eqnarray}
 where $X^{\nu}_{ii'}$ and $Y^{\nu}_{ii'}$ are forward and backward 
 amplitudes of the two-quasiparticle component $ii'$.
 Index $\nu$ stands for the QRPA normal mode whose
 excitation energy is denoted as $\hbar\omega_{\nu}$. For the time-reversal
 convention, we employ $Ta_{nljm}T^{\dagger}\equiv{\tilde a}_{nljm}
 =(-)^{l-j+m}a_{nlj-m}$.

 The $X$- and $Y$-amplitudes 
 are often obtained by diagonalization of the QRPA matrix\cite{Ring-Schuck}.
 Below, we shall show that one can also calculate the $X$- and
 $Y$-amplitudes on the basis of the linear response
 formalism.

 Let us consider transition densities 
\begin{eqnarray}
 \rho_{\alpha LM}^{({\rm tr})\nu}(\vecr)=
 \langle 0|{\hat \rho}_{\alpha}(\vecr)| \nu \rangle,
 \label{eq:tr0}
\end{eqnarray}
 for the three kinds of densities 
 ${\hat \rho}_\alpha(\vecr)={\hat \rho}(\vecr), 
  \hat{\tilde \rho}_{\pm}(\vecr)$.
 This quantity can be expressed in the following two ways. First, using
 the phonon creation operator $O^{\nu \dagger}_{LM}$ that describes the
 excited state$|\nu\rangle$ as $|\nu\rangle
 =O^{\nu\dagger}_{LM}|0\rangle$, Eq. (\ref{eq:tr0}) is written as
\begin{eqnarray}
 \rho_{\alpha LM}^{({\rm tr})\nu}(\vecr)
&=&\langle 0|\left[{\hat \rho}_{\alpha}(\vecr),O^{\nu\dagger}_{LM}\right]
| 0\rangle\\
\label{eq:TrP1}
&=&\sum_{i\geq i'}{1\over \sqrt{1+\delta_{ii'}}}\left\{
X_{ii'}^{\nu}\langle 0|\left[\hat{\rho}_{\alpha}(\vecr),\left[
a_{i}^{\dagger}a_{i'}^{\dagger}
\right]_{LM}\right]|0 \rangle
-Y_{ii'}^{\nu}\langle 0|\left[\hat{\rho}_{\alpha}(\vecr),\left[
a_{i'}^{ ~}a_{i}^{~ }
\right]_{LM}\right]|0 \rangle
\right\}\label{eq:trDef}
\\
&=&
Y_{LM}(\hat{\vecr}){1 \over r^2}
\sum_{i\geq i'}{1\over \sqrt{1+\delta_{ii'}}}
\left\{
(-)^{l'}X^{\nu}_{ii'}\frac{\langle l'j'||Y_{L}||lj\rangle^{*}}{\sqrt{2L+1}}
\langle 0|\rho_{\alpha}(r)|ii' \rangle
\right.\nonumber
\\
&~&~~~~~~~~~~~~~~~~~~~~~~~~~~~~~~~~~~~~~~~~~~~~
\left.
+(-)^{l} Y^{\nu}_{ii'}\frac{\langle l'j'||Y_{L}||lj\rangle^{*}}{\sqrt{2L+1}}
\langle ii'|\rho_{\alpha}(r)|0 \rangle
 \right\}.
\label{eq:TrP2}
\end{eqnarray}
 The other way is to use the density response
 $\delta\rho_{\alpha L}(r, \omega)$ of the linear response formalism.
 In the linear response formalism, the transition density can be
 calculated as
\begin{equation}
 \rho_{\alpha LM}^{({\rm tr})\nu}(\vecr)
= Y_{LM}(\hat{\vecr})\left[-\frac{C}{\pi r^2}{\rm Im}\delta\rho_{\alpha L}
(r,\omega_{\nu}) \right],
\end{equation}
 where $\omega_{\nu}$ is the frequency of the QRPA normal mode.
 We then rewrite this by using
 the linear response equation (Eq. (\ref{eq:drho})) and the
 unperturbed response function (Eq. (\ref{eq:spR0F})), and we obtain 
 an expression for the transition density 
\small
\begin{eqnarray}
 \rho_{\alpha LM}^{({\rm tr})\nu}(\vecr)
&=& Y_{LM}(\hat{ \vecr}){1\over r^{2}} \nonumber \\
&~&\times \sum_{i\geq i'}
\left\{
\left[
-\frac{C}{2\pi}
{\langle l'j'||Y_L||lj \rangle ^2 \over 2L+1}
\frac{1}{\hbar\omega_{\nu} -E_{i}-E_{i'}}
\right.\right. \nonumber
\\
&~&~~~~~~~~~~
\left.  \times
\int_0 d\rp \sum_{\beta}
\langle ii'|\rho_{\beta}(r')| 0\rangle
\left\{
\sum_{\gamma}\kappa_{\beta\gamma}(\rp){\rm Im}
\delta\rho_{\gamma L}(r',\omega_{\nu})
\right\}
\right] 
 \langle 0|\rho_{\alpha}(r)| ii'\rangle
 \nonumber \\
&~&
~~~~ +
\left[
\frac{C}{2\pi}
{\langle l'j'||Y_L||lj \rangle ^2 \over 2L+1}
\frac{1}{\hbar \omega_{\nu} +E_{i}+E_{i'}}\right. \nonumber
\\
&~&~~~~~~~~
\left.\left.  \times
\int_0 d\rp \sum_{\beta}
\langle 0|\rho_{\beta}(r')| ii'\rangle
\left\{
\sum_{\gamma}\kappa_{\beta\gamma}(\rp){\rm Im}
\delta\rho_{\gamma L}(r',\omega_{\nu})
\right\}
\right] \langle ii'|\rho_{\alpha}(r)| 0\rangle
\right\}.
\label{eq:Trl}
\end{eqnarray}
\normalsize
 Note that since the density response 
 $\delta\rho_{\alpha L}(r, \omega)$ exhibits a pole behavior
 $\propto 1/(\hbar\omega + i\eps - \hbar\omega_\nu)$ at frequencies 
 near the QRPA eigen mode $\hbar\omega_\nu$, the external
 field $v^{ext}_{\beta L}(r)$ in Eq. (\ref{eq:drho}) is irrelevant here,
 provided
 that we take a sufficiently small value of $\epsilon$. The
 coefficient $C$ is a normalization constant.

 Comparing the above two definitions of the transition density,
 Eq. (\ref{eq:TrP2}) and Eq. (\ref{eq:Trl}),
 we obtain the expression for 
 the forward and backward amplitudes $X_{ii'}^{\nu}$ and $Y_{ii'}^{\nu}$,
\begin{eqnarray}
X^{\nu}_{ii'}
&=&
(-)^{l'+1}
\frac{C}{2\pi}
\frac{\langle l'j'||Y_L||lj \rangle}{\sqrt{2L+1}}
\frac{1}
{\hbar\omega_{\nu} -E_{i}-E_{i'}}\nonumber\\
&~&~~~~~~~~~~~~~~~~~~~~~~\times
\int_0 d\rp \sum_{\beta}
\langle ii'|\rho_{\beta}(r')| 0\rangle
\left\{
\sum_{\gamma}
\frac{\partial v_{\beta}}{\partial \rho_{\gamma}}(r'){1 \over \rp ^{2}}
{\rm Im}\delta\rho_{\gamma L}(r',\omega_{\nu})
\right\},\label{eq:xij} \\
Y^{\nu}_{ii'}
&=&
(-)^{l}
\frac{C}{2\pi}
\frac{\langle l'j'||Y_L||lj \rangle}{\sqrt{2L+1}}
\frac{1}
{\hbar \omega_{\nu} +E_{i}+E_{i'}}\nonumber\\
&~&~~~~~~~~~~~~~~~~~~~~~~\times
\int_0 d\rp \sum_{\beta}
\langle 0|\rho_{\beta}(r')| ii'\rangle
\left\{
\sum_{\gamma}
\frac{\partial v_{\beta}}{\partial \rho_{\gamma}}(r'){1 \over \rp ^{2}}
{\rm Im}\delta\rho_{\gamma L}(r',\omega_{\nu})
\right\}.\label{eq:yij}
\end{eqnarray}
 The normalization constant $C$ is determined to fulfill the normalization condition
$\langle 0|\left[O^{\nu}_{LM},O^{\nu\dagger}_{LM} \right]|0\rangle 
= \sum_{i\geq i'}|X_{ii'}^\nu|^2-|Y_{ii'}^\nu|^2 =1$.

\subsection{Decomposition of the transition densities}

 Once the $X$- and $Y$-amplitudes are obtained, it is
 straightforward to 
 decompose the transition density with respect to
 the two-quasiparticle components of the QRPA phonon:
\begin{eqnarray}
 \rho^{({\rm tr})\nu}_{\alpha LM}(\vecr)
&=&\sum_{i\geq i'}\rho^{({\rm tr})\nu}_{\alpha LM,ii'}(\vecr),
\label{eq:TrP4a} \\
\rho^{({\rm tr})\nu}_{\alpha LM,ii'}(\vecr)\label{eq:TrP4b}
&=&{1\over \sqrt{1+\delta_{ii'}}}\left\{
X_{ii'}^{\nu}\langle 0|\left[\hat{\rho}_{\alpha}(\vecr),\left[
a_{i}^{\dagger}a_{i'}^{\dagger}
\right]_{LM}\right]|0 \rangle
-Y_{ii'}^{\nu}\langle 0|\left[\hat{\rho}_{\alpha}(\vecr),\left[
a_{i'}^{ ~}a_{i}^{~ }
\right]_{LM}\right]|0 \rangle\right\}.\label{eq:TrP4c}
 \end{eqnarray}
 Here $\rho^{({\rm tr})\nu}_{\alpha LM,ii'}(\vecr)$ is a partial
 contribution from a two-quasiparticle configuration $ii'$. Its radial
 part is expressed as 
\begin{eqnarray}
 \rho_{\alpha LM,ii'}^{({\rm tr})\nu}(\vecr)&=&
Y_{LM}(\hat{\vecr}) \rho_{\alpha L,ii'}^{({\rm tr})\nu}(r),
\label{eq:TrP3a}\\
 \rho_{\alpha L,ii'}^{({\rm tr})\nu}(r)
&=&\frac{\langle l'j'||Y_{L}||lj\rangle^{*}}{\sqrt{2L+1}}{1 \over r^2}
\left\{
(-)^{l'}X^{\nu}_{ii'}\langle 0|\rho_{\alpha}(r)|ii' \rangle
+(-)^{l} Y^{\nu}_{ii'}\langle ii'|\rho_{\alpha}(r)|0 \rangle
 \right\}
\label{eq:TrP3b}.
\end{eqnarray}

 In the present paper, we describe the pair transfer modes 
 in terms of operators
\begin{eqnarray}
P^{\dagger}(\vecr)
&=&\psi^{\dagger}(\vecr \downarrow)\psi^{\dagger}(\vecr\uparrow)
={1\over 2}\left(
\hat{\tilde \rho}_{+}(\vecr)-\hat{\tilde \rho}_{-}(\vecr)
\right),\\
P^{~}(\vecr)
&=&\psi(\vecr\uparrow)\psi(\vecr\downarrow)
= {1\over 2}\left(
\hat{\tilde \rho}_{+}(\vecr)+\hat{\tilde \rho}_{-}(\vecr)
\right),
\end{eqnarray} 
 which adds and removes a spin-singlet ($S$=0) neutron pair,
 respectively.
 We then describe transition densities for
 these pair-addition and pair-removal operators, defined by
\begin{eqnarray}
P_{\nu LM}^{({\rm ad})}(\vecr)&\equiv&
\langle 0 |
\psi(\vecr\downarrow)\psi(\vecr\uparrow)
|\nu \rangle
=P_{\nu L}^{({\rm ad})}(r)Y_{LM}(\hat{\vecr}),
\\
P_{\nu LM}^{({\rm rm})}(\vecr)&\equiv&
\langle 0 |
\psi^{\dagger}(\vecr\downarrow)\psi^{\dagger}(\vecr\uparrow)
|\nu \rangle 
=P_{\nu L}^{({\rm rm})}(r)Y_{LM}(\hat{\vecr}),
\end{eqnarray}
 which are decomposed as
\begin{equation}
P_{\nu L}^{({\rm ad/rm})}(r)= \sum_{i \geq i'}
P_{\nu L,ii'}^{({\rm ad/rm})}(r),
\end{equation}
 and
\begin{eqnarray}
P_{\nu L,ii'}^{({\rm ad/rm})}(r)
&=&{1\over \sqrt{1+\delta_{ii'}}}
\frac{\langle l'j'||Y_{L}||lj\rangle^{*}}{2r^{2}}
\left[
(-)^{l'}X^{\nu}_{ii'}
\left\{
\langle 0|\tilde{\rho_{+}}(r)|ii' \rangle
\pm\langle 0|\tilde{\rho_{-}}(r)|ii' \rangle
\right\}\right.\nonumber \\ 
&~&~~~~~~~~~~~~~~~~~~~~~~~~~~~~~~~~~~
+(-)^{l} Y^{\nu}_{ii'}\left.
\left\{
\langle ii'|\tilde{\rho_{+}}(r)|0 \rangle
\pm\langle ii'|\tilde{\rho_{-}}(r)|0 \rangle
\right\}
 \right].
\label{eq:TrP3aii}
\end{eqnarray}

 We can also consider the transition density of
 the pair-rotational ground-state transfer\cite{Shimoyama11}.
 The transition density is
 also expanded using the quasiparticle states:
\begin{eqnarray}
P^{({\rm ad/rm})}_{\rm gs}(\vecr)
&=&
\langle 0|
\psi^{\dagger}(\vecr\sigma)\psi^{\dagger}(\vecr{\tilde\sigma})
|0\rangle \label{eq:TrDgs1}\\
&=&
-\langle 0|
\sum_{ii'}\sum_{mm'}
\varphi^{~}_{2,i}(\vecr{\tilde \sigma})
\varphi^{*}_{1,i'}(\vecr{\tilde\sigma})
a^{~}_{i}a^{\dagger}_{i'}
|0\rangle \label{eq:TrDgs2}\\
&=&
-\frac{1}{4\sqrt{\pi}}{1\over r^2}
\sum_{i}
\varphi^{~}_{2,i}(r)
\varphi^{*}_{1,i}(r)Y_{00}(\hat{\vecr}).\label{eq:TrDgs3}
\end{eqnarray}
Then, the radial transition density of the ground state transfer and its
decomposition are 
\begin{eqnarray}
P^{({\rm ad/rm})}_{\rm gs}(r)&=& \sum_i 
P^{({\rm ad/rm})}_{{\rm gs},ii}(r),
\label{eq:TrDgs4}\\
P^{(\rm ad/rm)}_{{\rm gs},ii}(r)&=&
-\frac{1}{4\sqrt{\pi}}{1\over r^2}(2j+1)
\varphi^{~}_{2,i}(r)
\varphi_{1,i}(r).
\label{eq:TrDgs5}
\end{eqnarray}

\subsection{Details of calculation}
 We assume the spherical ground states for the proton
 closed-shell nuclei Sn ($Z$=50).
 The Skyrme interaction parameter SLy4 \cite{chabanat98} is chosen for the
 particle-hole channel of the HFB equations. The Landau-Migdal
 approximation is employed for the RPA residual interaction in the
 particle-hole channel. For the pairing interaction we adopt
 density-dependent delta interaction (DDDI)
 \cite{Esbessen92,Dobaczewski01,Chasman76,Garrido01}.
 The DDDI is given by
 $v_{q}^{\rm pair}(\vecr,\vecrp)={1\over
 2}V_{q}(\vecr)\delta(\vecr-\vecrp)$, where $V_{q}(\vecr)$ is the
 pairing interaction strength and is a function of the neutron and proton
 densities. We adopt the following form: 
\begin{eqnarray}
V_{q}(\vecr)=v_{0}\left[1-\eta
\left(\frac{\rho_q(\vecr)}{\rho_0}\right)^{\alpha}\right].
\label{eq:dddi}
\end{eqnarray}
 The value $v_0$ = -458.4 MeV fm$^3$ is chosen to reproduce the scattering
 length $a$ = -18.5 fm of the bare neutron-neutron interaction in the
 $^{1}S$ channel.
 The parameter $\eta$=0.71 is adjusted to reproduce the experimental
 pairing gap in $^{120}$Sn.
 The other parameters are $\rho_0= 0.08$ fm$^{-3}$ and $\alpha= 0.59$
 \cite{Satula1998,Matsuo06}.

 We solve the HFB equation in the coordinate space representation using
 the polar coordinate system. The radial coordinate space is truncated at
 $r_{\rm max}$= 20 fm.
 We truncate the quasiparticle states by putting
 the maximum quasiparticle energy $E_{\rm max}$= 60 MeV and the
 largest value of the orbital angular quantum number $l_{\rm max} = 12$.
 The quasiparticle states in the continuum energy
 region are obtained with the box boundary condition
 $\phi(r_{\rm max})=0$, and they are all discretized.
 All other details can be found in Ref.\cite{Matsuo10,Shimoyama11}.

 We calculate the strength function
$ S_{{\rm Pad}0}(\hbar\omega) \equiv 
 \sum_{\nu} |\langle \nu
 |P^{\dagger}_{00}|0\rangle|^2
\delta(\hbar\omega-\hbar\omega_{\nu})$
 for the monopole ($L$=0)
 pair-addition transfer and the strength function
 $ S_{{\rm Prm}0}(\hbar\omega)
 \equiv 
 \sum_{\nu} |\langle \nu
 |P^{~}_{00}|0 \rangle|^2
\delta(\hbar\omega-\hbar\omega_{\nu})$
 for the pair-removal transfer,
 where, the pair-addition operator $P^{\dagger}_{LM}$
 and pair-removal operator $P_{LM}$,
\begin{eqnarray}
P^\dagger_{LM} &=& \int d\vecr Y_{LM}(\hat{r})
\psi^\dagger(\vecr\downarrow)\psi^\dagger(\vecr\uparrow),\\
P_{LM} &=& \int d\vecr Y_{LM}(\hat{r})
\psi(\vecr\uparrow)\psi(\vecr\downarrow),
\end{eqnarray}
 are creation and annihilation operators of the $S$ = 0 neutron pair
 with the total angular momentum $L$. 
 These strength functions are evaluated 
 using the solutions of the linear response equation
 (Eq. (\ref{eq:drho}))\cite{Matsuo10}.
 Concerning the small imaginary part $i\eps$ appearing in the  
 response function (Eq. (\ref{eq:spR0F})), we choose $\eps=0.5$ keV, which is 
 much smaller than the value $\eps=50$ keV adopted in Ref.\cite{Shimoyama11}.
 The QRPA eigen energies $\hbar\omega_\nu$ are then extracted by searching the
 peak energies of the pair-addition 
 strength function $S_{{\rm Pad}0}(\hbar\omega)$.
 The pair-addition strength 
\begin{equation}
B({\rm Pad}0;{\rm gs}\rightarrow \nu)\equiv
|\langle 0 |P^{~}_{00}|\nu \rangle|^{2}
\end{equation}
 associated with individual QRPA eigen modes is evaluated by integrating
 the strength function $S_{{\rm Pad}0}(\hbar\omega)$ in an energy 
 interval $\in \left[\hbar\omega_{\nu}-10\eps,\hbar\omega_{\nu}+10\eps\right]$.
 We also calculate the strength $B({\rm Pad}0;{\rm gs}\rightarrow {\rm gs})$
 of the ground-state transfer (the pair rotation) as
 $B({\rm Pad}0;{\rm gs}\rightarrow {\rm gs})
=\left|\int r^{2}P^{({\rm ad})}_{\rm gs}(r)dr\right|^2$
 using the transition density $P^{({\rm ad})}_{\rm gs}(r)$.

\section{Various collective pair-transfer modes in Sn isotopes}

\begin{figure}[tp] 
\includegraphics[width=7.9cm]{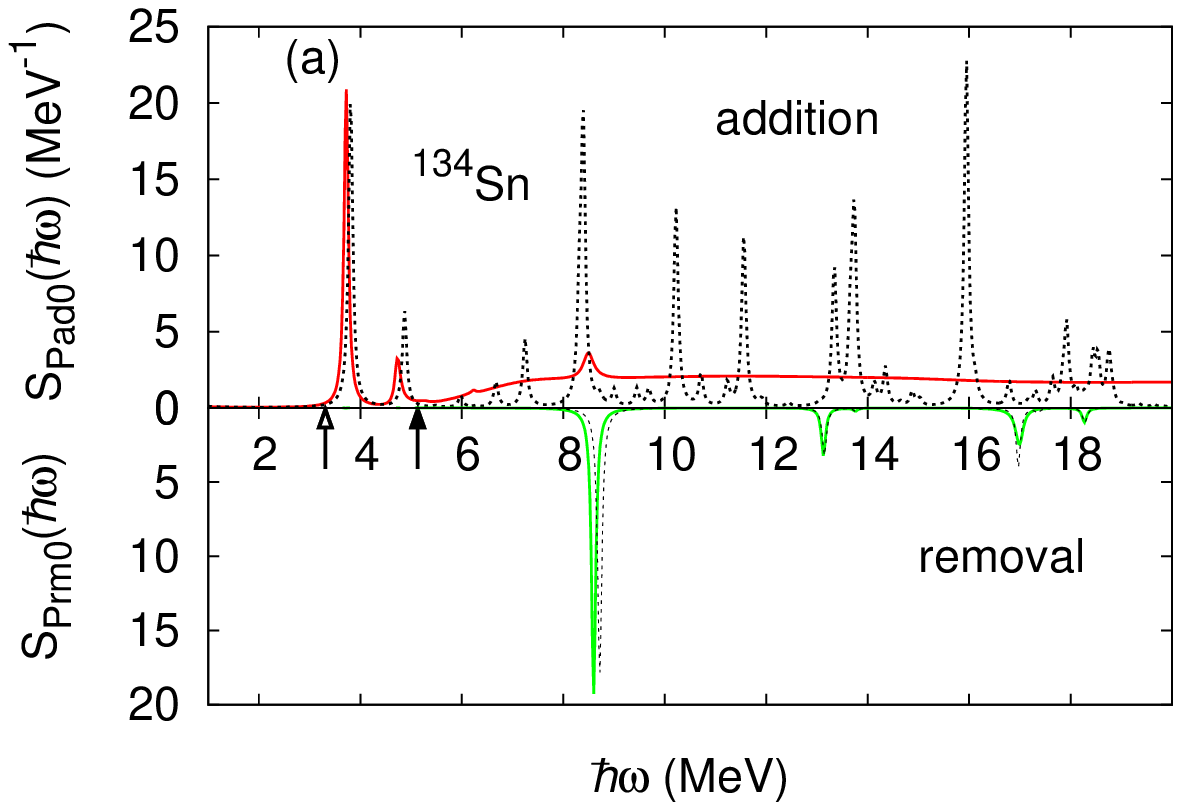}
\includegraphics[width=7.9cm]{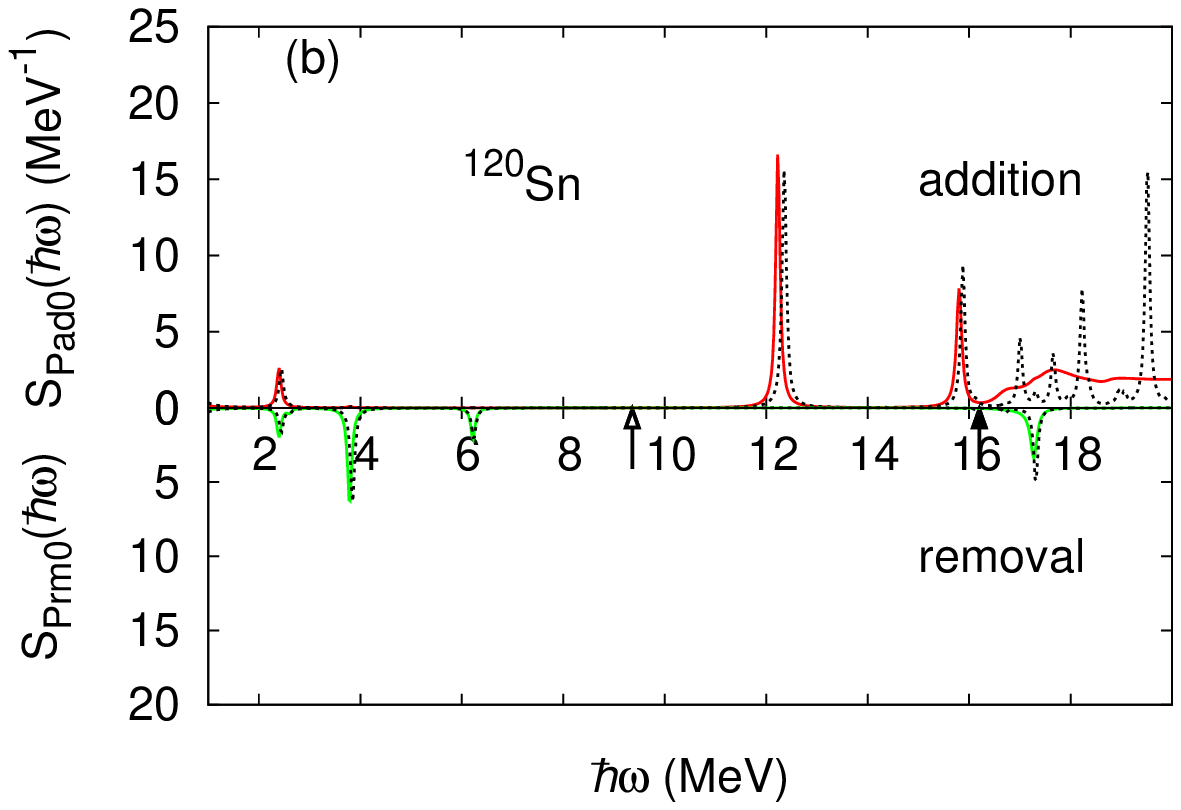}
\caption{\label{fig:SF-gpv}(Color online)
 The strength function $S_{\rm Pad0}(\hbar\omega)$ 
 for the pair-addition mode,
 plotted in the upper panel,
 and the strength function $S_{\rm Prm0}(\hbar\omega)$
 for the pair-removal mode (in the lower panel)
 in (a) $^{134}$Sn and (b) $^{120}$Sn. The dotted curves are
 the results of the discretized QRPA while
the solid curves are the result of the continuum QRPA calculation.
The smoothing parameter $\eps$= 50 keV is adopted in this figure.
 The arrows indicate the one- and the two-neutron separation energies
 $S_{1n}$=3.31 MeV and $S_{2n}$=5.13 MeV for $^{134}$Sn,
 $S_{1n}$=9.36 MeV and $S_{2n}$=16.20 MeV for $^{120}$Sn.
}
\end{figure}

 Figures \ref{fig:SF-gpv}(a) and \ref{fig:SF-gpv}(b) show the strength
 functions
 $S_{{\rm Pad}0}(\hbar\omega)$ and $S_{{\rm Prm}0}(\hbar\omega)$
 for the monopole ($L$=0)
 pair-addition and pair-removal transfers in neutron-rich $^{134}$Sn
 and stable $^{120}$Sn, respectively.
 In Fig. \ref{fig:SF-gpv} we also show results of the 
 continuum QRPA calculation with solid curves.
 In this figure and only here, we used a slightly 
 large value of the imaginary constant $\eps$= 50 keV in order to 
 make the peaks visible. It is noted here that most of the peaks
 seen in the pair-addition strength function
 $S_{{\rm Pad}0}(\hbar\omega)$ in high energy region above
 the two neutron separation energy (the second arrow) are
 fictitious peaks originating from the discretization. We shall
 focus only on peaks for which 
 the discretized and the continuum calculations give
 essentially the same peak energies and strengths.

 First, we pay attention to the low-lying peaks at
 2$\leq\hbar\omega\leq$4 MeV. As already discussed in
 Ref.\cite{Shimoyama11}, a prominent feature 
 in neutron-rich $^{134}$Sn is
 the large peak located at $\hbar\omega\approx 3.8$ MeV in 
 the pair-addition strength function 
 $S_{{\rm Pad}0}(\hbar\omega)$.
 The pair-addition strength of this pair vibration
 is several times larger than that of the low-lying
 pair vibration at $\hbar\omega\approx 2.4$ MeV
 in stable $^{120}$Sn. 

 Let us next look at higher-lying pair-transfer modes up to 
 $\hbar\omega= 20$ MeV, extending our previous study
 which covered only $\hbar\omega< 10$ MeV.
 There is no significant peak in the pair-addition strength
 $ S_{{\rm Pad}0}(\hbar\omega)$ in the 
 high-frequency region in neutron-rich $^{134}$Sn except
 the smooth strength distribution spread broadly above the
 two neutron separation energy. 
 In stable $^{120}$Sn, on the other hand,
 two modes with large pair-addition strength
 appear at $\hbar\omega\approx 12.4$ MeV and
 $\hbar\omega\approx 15.9$ MeV. They are located below the
 two-neutron separation energy $S_{2n}=16.2$ MeV. In the continuum QRPA
 calculation,  these peaks have physical escaping widths since they are
 located above the one-neutron separation  energy $S_{1n}=9.36$ MeV, but
 their widths are actually very small.  As we show later, these two
 peaks can be regarded as so called giant pair vibration
 (GPV)\cite{Oertzen2011,Broglia1977,Lotti1989}.  The GPV is a collective
 pair-transfer mode of  adding or removing two neutrons in the next
 major shell beyond the valence shell. In the present calculation, the giant
 pair vibration consists of two peaks.  We call these two peaks GPV1
 ($\hbar\omega\approx 12.4$ MeV) and GPV2 ($\hbar\omega\approx 15.9$
 MeV), and we discuss them in detail in Section \ref{sec:gpv}.  The
 giant pair vibrations exist in all the isotopes with $110\lesim
 A<132$.

 In the pair-removal strength function 
 $ S_{{\rm Prm}0}(\hbar\omega)$ in $^{134}$Sn (Fig. \ref{fig:SF-gpv}(a))
 there exist a giant pair-removal vibration with large strength 
 at $\hbar\omega\approx 8.7$ MeV. 
 It corresponds to the neutron pair-removal from the
 $N= 50-82$ shell\cite{Herzog85}.
 In this paper, however, 
 we do not investigate the pair-removal modes.\\

\begin{figure}[tp]
\includegraphics[width=7.9cm]{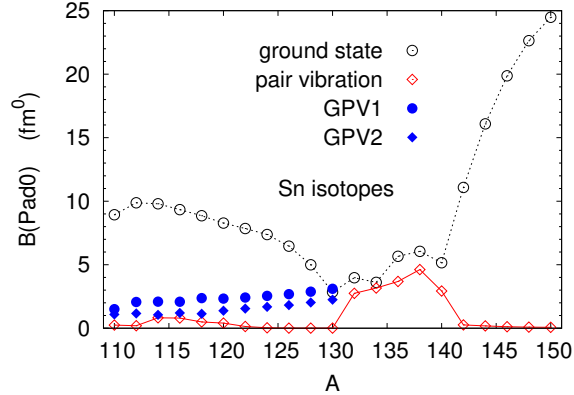}
\caption{\label{fig:st-Sn}(Color online)
 The neutron pair-addition strength 
 $B({\rm Pad}0;{\rm gs}\rightarrow \nu)$ for
 the giant pair vibrational modes GPV1
 and GPV2, plotted with the filled circle and diamond, respectively,
 and the same quantity for the low-lying pair pair vibrational
 mode, plotted with open diamond, as well as
 the neutron pair-addition strength $B({\rm Pad}0; {\rm gs} \rightarrow {\rm gs})$
 for the pair-rotational ground-state transfer (open circle), in
 even-even Sn isotopes.
 }
\end{figure}
 
 In Fig. \ref{fig:st-Sn},
 we show the systematical behavior of the pair-additional strength
 $B({\rm Pad}0;{\rm gs}\rightarrow \nu)$ for 
 the two giant pair vibrations GPV1 and GPV2,
 as well as the strength $B({\rm Pad}0;{\rm gs}\rightarrow \nu)$
 of the low-lying pair vibration,
 and 
 the strength $B({\rm Pad}0;{\rm gs}\rightarrow {\rm gs})$
 for the pair-rotational ground-state transfer in the Sn isotopes $A=110-150$.
 It is seen that the pair-addition strengths of GPV1 and GPV2 increase
 with the neutron number, and at $A\sim 130$
 they become comparable to those of the characteristic low-lying pair
 vibrations and those of the ground-state transfer in $^{132-140}$Sn.

 The values of the pair-addition strength and the excitation energy
 for the giant pair vibrational states are also listed
 in Table \ref{tabl:Sngpv2X} and \ref{tabl:Sngpv1X}.

\section{Microscopic structures 
of low-lying pair vibrational modes \label{sec:low-lying}}

\subsection{Dominant two-quasiparticle configurations}

 In Ref.\cite{Shimoyama11},
 we have discussed the collectivity
 of the low-lying pair-addition vibration ($\hbar\omega_{\nu}=3.81$ MeV)
 in $^{134}$Sn by comparing its strength $B({\rm Pad}0)=3.23$ 
 with the single-particle strength $B_{s.p.}({\rm Pad}0)=(2j+1)/8\pi=$
 0.18 and 0.15, evaluated for pure independent two-neutron configurations
 $[2f_{7/2}]^{2}$ and $[3p_{3/2}]^{2}$, respectively. We observe
 a large collective enhancement of a factor of more than ten.
 In this section, we reveal the origin of 
 this collectivity by examining 
 microscopic structure of this mode. 

 First we
 evaluate and analyze
 the forward and backward amplitudes $X_{ii'}^{\nu}$ and
 $Y_{ii'}^{\nu}$ 
 for the two-quasiparticle configurations, given by
 Eqs.(\ref{eq:xij}) and (\ref{eq:yij}), respectively.
 We find that the two-neutron transfer modes discussed above almost all
 the amplitudes ($\gtsim 99\%$) are exhausted by neutron two-quasiparticle
 states. In the following, we discuss only the neutron amplitudes.

\begin{table}[tp]
\begin{center}
\begin{tabular}{cccrr}
 \hline\hline
&&$E_{i}+E_{i'}$&\multicolumn{1}{c}{$X^{\nu}_{ii'}$}&\multicolumn{1}{c}{$Y^{\nu}_{ii'}$} \\
 \hline
$^{134}$Sn
&$[3p_{3/2}]^{2}$     & 4.49 & 0.752 & 0.0006 \\
$\hbar\omega_{\nu}$=3.81 MeV
&$[1h_{9/2}]^{2}$     & 6.71 & -0.368 & -0.0017 \\
&$[2f_{7/2}]^{2}$     & 1.50 & -0.297 & -0.0325 \\
&$[2f_{5/2}]^{2}$     & 6.35 & -0.279 & -0.0013 \\
&$[1i_{13/2}]^{2}$    & 9.61 & 0.241 & 0.0021 \\
&$[3p_{1/2}]^{2}$     & 5.52 & 0.177 & -0.0005 \\
&$[3p_{3/2}][4p_{3/2}]$  & 6.78 & 0.101 & -0.0004 \\
 \hline
$^{120}$Sn
&$[1h_{11/2}]^{2}$ & 3.60 & 0.659 & -0.059 \\
$\hbar\omega_{\nu}$= 2.44 MeV
&$[2d_{ 3/2}]^{2}$ & 2.52 & 0.606 & 0.014 \\
&$[3s_{ 1/2}]^{2}$ & 2.85 & -0.436 & -0.021 \\
&$[1g_{ 7/2}]^{2}$ & 4.80 & 0.087 & -0.072 \\
&$[2d_{ 5/2}]^{2}$ & 6.76 & -0.079 & 0.039 \\
 \hline\hline
\end{tabular}
  \caption{\label{tabl:sn120134pvXY}
 The forward amplitude $X^{\nu}_{ii'}$ and
 the backward amplitude $Y^{\nu}_{ii'}$
 for the low-lying pair vibration modes 
 in $^{134}$Sn ($\hbar\omega_{\nu}= 3.81$ MeV) and 
 in $^{120}$Sn ($\hbar\omega_{\nu}= 2.44$ MeV).
 The two-quasiparticle configurations which have the largest
 values of $|X_{ii'}^{\nu}|$ are listed. 
 }
\end{center}
\end{table}

\begin{figure}[tp]
\includegraphics[width=7.9cm]{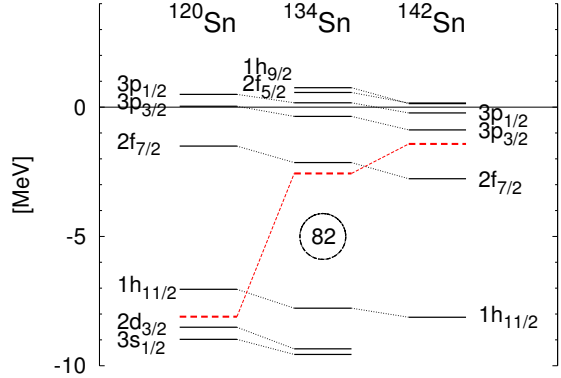}
\caption{\label{fig:spe}(Color online)
 The Hartree-Fock single-particle energies $e_{\rm HF}$ of neutrons
 in $^{120}$Sn, $^{134}$Sn
 and $^{142}$Sn.
 The dashed lines show the neutron Fermi energies
 $\lambda_n = -8.10, -2.56$ and $-1.42$ MeV for the three 
 isotopes.}
\end{figure}

 Let us consider the pair vibration at $\hbar\omega_{\nu}= 3.81$ MeV
 in neutron-rich $^{134}$Sn.
 Table \ref{tabl:sn120134pvXY} lists the two-quasiparticle components
 having the largest absolute values of the 
 forward amplitude $|X_{ii'}^{\nu}|$.
 The component with the largest amplitude is
 $[3p_{3/2}]^{2}$ with $X_{ii'}^{\nu}=0.752$.
 There are seven components whose $X$-amplitudes satisfy
 $|X_{ii'}^{\nu}|>0.1$; $[3p_{3/2}]^{2}$, $[1h_{9/2}]^{2}$,
 $[2f_{7/2}]^{2}$, $[2f_{5/2}]^{2}$ $[1i_{13/2}]^{2}$ and
 $[3p_{1/2}]^{2}$.
 The main components indicate that 
 the low-lying pair vibrational state in $^{134}$Sn
 has some degree of collectivity.
 The Hartree-Fock(HF) single-particle energies of neutrons 
 are shown in Fig. \ref{fig:spe}.
 The quasiparticle orbits of the main components
 are those located near the Fermi energy, all of which are either
 weakly bound or unbound resonant quasiparticle states.

 In the case of the low-lying pair vibration 
 ($\hbar\omega_{\nu}$= 2.44 MeV) in stable $^{120}$Sn,
 there are only three components
 $[1h_{11/2}]^{2}$, $[2d_{3/2}]^{2}$ and $[3s_{1/2}]^{2}$, having large
 forward amplitudes satisfying $|X_{ii'}^{\nu}|>0.1$, see Table
 \ref{tabl:sn120134pvXY}. 
 It is evident that the collectivity is smaller here than
 that of the pair vibration in $^{134}$Sn.
 This is consistent qualitatively with a much smaller value of
 the pair-addition strength $B({\rm Pad}0)=0.40$
 in $^{120}$Sn than $B({\rm Pad}0)=3.16$ in $^{134}$Sn.

\begin{figure}[tp]
\includegraphics[width=7.9cm]{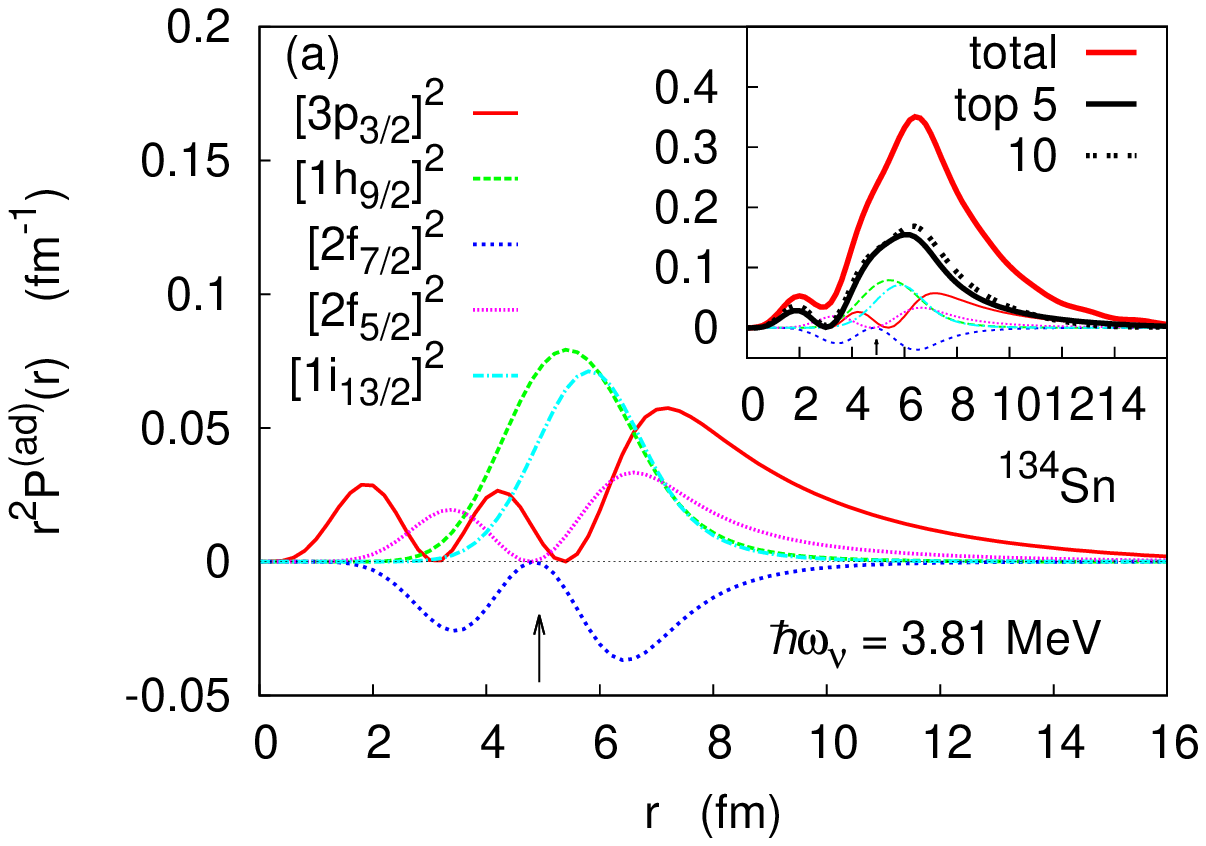}
\includegraphics[width=7.9cm]{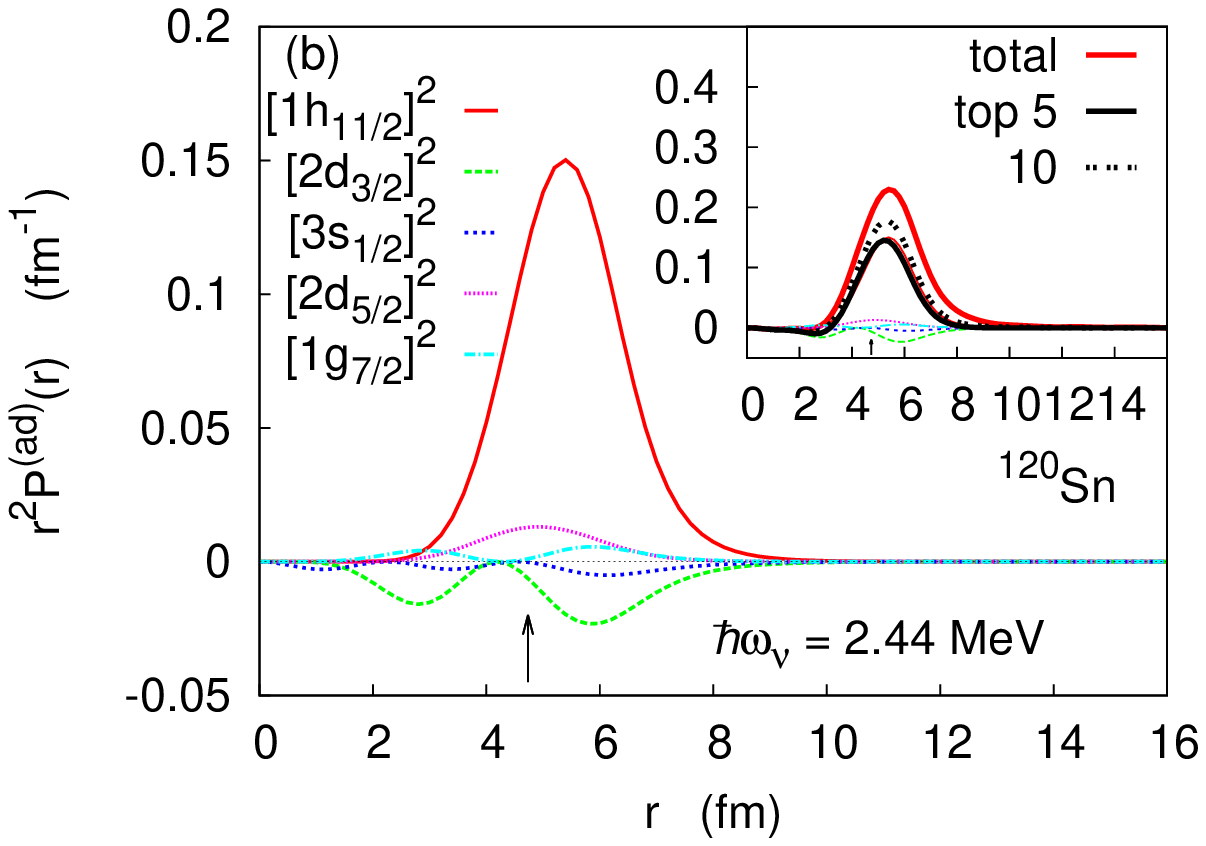}
\caption{\label{fig:TrD-Sn120134top}(Color online)
 (a) The decomposed transition densities 
 $r^{2}P^{({\rm ad})}_{\nu L,ii'}(r)$ 
 for the largest two-quasiparticle components of the 
 low-lying pair vibration mode ($\hbar\omega_{\nu}=3.81$ MeV)
 in $^{134}$Sn (cf. Table \ref{tabl:sn120134pvXY}). 
 In the inset, shown also are 
 the total transition density
 $r^{2}P^{({\rm ad})}_{\nu L}(r)$ and partial sums
 of the decomposed transition densities of the largest 
 five and ten two-quasiparticle components. 
 (b) The same as (a), but for the low-lying pair
 vibration mode ($\hbar\omega_{\nu}=2.44$ MeV)
 in $^{120}$Sn.
 The arrow indicates the neutron rms radius
 $R_{N,rms}(=\sqrt{\langle r^{2}_{n}\rangle})= 4.93$ fm and $4.73$ fm
 for $^{134}$Sn and $^{120}$Sn, respectively.
 }
\end{figure}

 Next, we analyze the transition density 
 $P_{\nu L}^{({\rm ad})}(r)$ by looking into 
 the decomposed transition densities 
 $P_{\nu L,ii'}^{({\rm ad})}(r)$ (Eq.(\ref{eq:TrP3aii}))
 associated with the main two-quasiparticle configurations.
 They are shown in 
 Figs. \ref{fig:TrD-Sn120134top}(a) and \ref{fig:TrD-Sn120134top}(b),
 for the low-lying pair vibration in $^{134}$Sn and $^{120}$Sn,
 respectively.

 We see in Fig. \ref{fig:TrD-Sn120134top}(a) 
 the following characteristics of the pair vibration in $^{134}$Sn.
 The individual decomposed transition densities have
 much smaller amplitudes than that of the total transition density.
 Even if we sum the transition densities of the largest five components
 (the thick solid curve in the inset), it accounts for only approximately
 half of the total transition density. The situation is essentially the
 same also for a sum of the largest ten components (the thick dotted curve). 
 We note also that 
 the tail of the total transition density extending to $r\sim16$ fm
 is not reproduced by the largest five (ten) components. 
 The component $[3p_{3/2}]^{2}$ has
 the largest tail among the five (note that $3p_{3/2}$ is weakly
 bound single-particle orbit with low angular momentum), but its 
 tail amplitude is quite smaller than that of total transition density. 
 We thus find that 
 contributions from small components other than the largest five or ten
 play important role in some way to produce the large pair-addition strength
 $B({\rm Pad}0)$ and the characteristic transition density of 
 the low-lying pair vibration in $^{134}$Sn.

 The situation of the low-lying pair vibration in
 $^{120}$Sn is different. As shown in
 Fig. \ref{fig:TrD-Sn120134top}(b), 
 the $[1h_{11/2}]^{2}$ component 
 has a dominant contribution to the pair-addition transition density,
 amounting 
 more than two-thirds of the total transition density around the surface.
 Because the HF single-particle energy of the 1$h_{11/2}$ orbit
 in $^{120}$Sn is located deeply at $e_{1h_{11/2}}=-7.00$ MeV,
 and the Fermi energy is also deep $\lambda_n= -8.10$ MeV
 (cf. Fig. \ref{fig:spe}),
 the transition density of the component $[1h_{11/2}]^{2}$
 as well as the total do not extend far outside the surface.

\subsection{High-$l$ two-quasiparticle configurations and 
di-neutron correlation}

 We have seen in the previous section that
 the main components of the pair vibration mode in 
 $^{134}$Sn reproduces only about a half of the total transition density. 
 In this subsection 
 we investigate how other small two-quasiparticle components contribute.

\begin{figure}[tp]
\includegraphics[width=7.9cm]{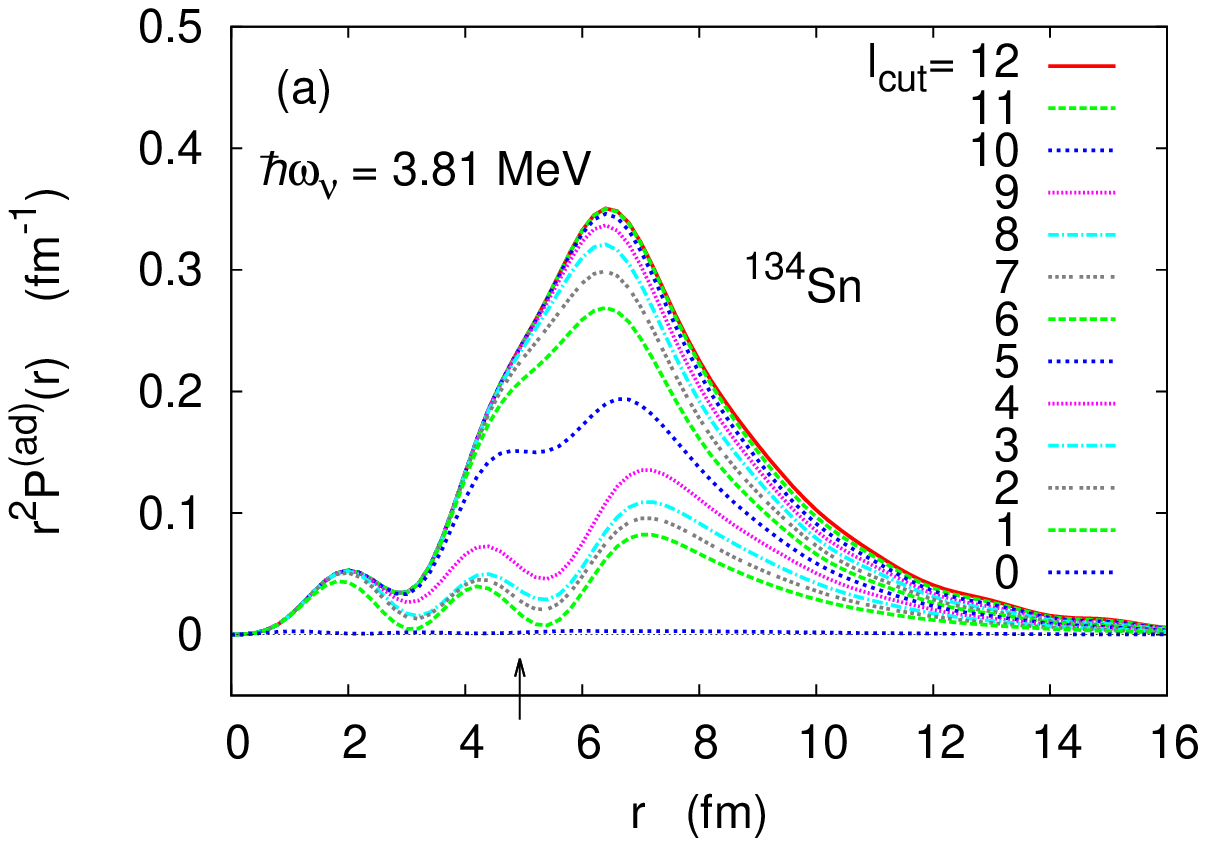}
\includegraphics[width=7.9cm]{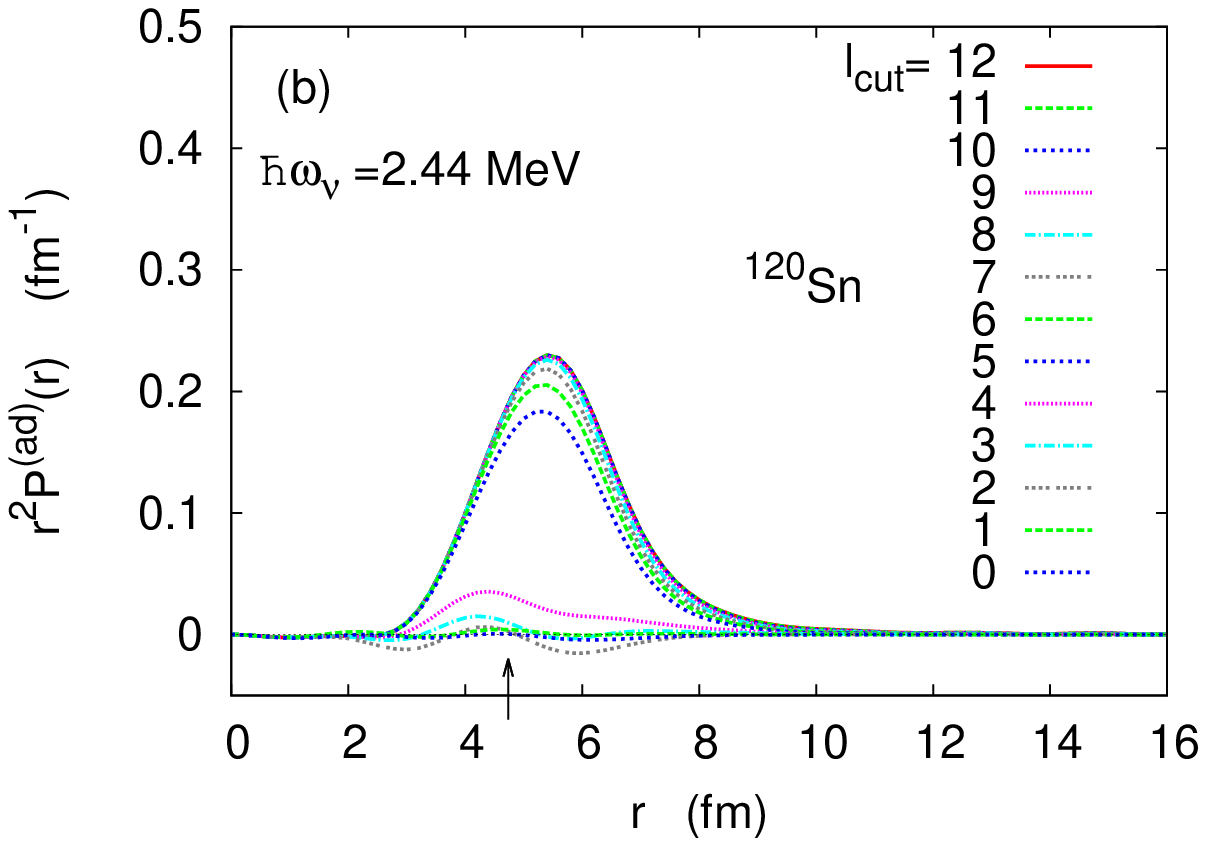}
\caption{\label{fig:TrD-lcutpv}(Color online)
 The $l$-cutted transition densities
 $r^{2}P^{({\rm ad})}_{\nu L,l_{cut}}(r)$
 with $l_{cut}$=0, 1, 2, $\cdots$,12 
 for the low-lying pair vibration modes
 (a) in $^{134}$Sn and (b) in $^{120}$Sn.
The arrow indicates the neutron rms radius $R_{N,rms}$.}
\end{figure}

 For this purpose, we put all the two-quasiparticle configurations
 into subgroups which are specified with the orbital angular momentum
 $l$ of the quasiparticle states. Note that we have
 only
 two-quasiparticle combinations $ii'=[nlj][n'l'j']$ satisfying $l=l'$ and
 $j=j'$ for $0^+$ modes. We then calculate a partial sum of
 the decomposed transition densities
 $P^{({\rm ad})}_{\nu L,l_{cut}}(r) =
 \sum_{ii',l\le l_{cut}}P^{({\rm ad})}_{\nu L,ii'}(r)$,
 where the orbital angular momenta $l$ are taken into account up to
 a cut-off values $l_{cut}$.
 Figures \ref{fig:TrD-lcutpv}(a) and \ref{fig:TrD-lcutpv}(b) 
 show the $l$-cutted transition densities
 $P^{({\rm ad})}_{\nu L,l_{cut}}(r)$ with 
 $l_{cut}=0, 1, 2, \cdots, 12$ 
 for low-lying pair-addition vibration
 in neutron-rich $^{134}$Sn and in stable $^{120}$Sn,
 respectively.

 Let us first look at the case of $^{134}$Sn, i.e. 
 Fig. \ref{fig:TrD-lcutpv}(a). It is clearly seen that
 a large number of the orbital angular momenta covering
 up to the maximum value $l_{max}$=12 have non-negligible
 contributions, especially at larger values of $r$.
 The highest orbital angular momentum of the occupied HF single-particle
 orbits in $^{134}$Sn is $l_{occ}= 5$ (1$h_{11/2}$).
 The sum up to $l_{cut}=l_{occ}= 5$ can account for only about a
 half of the total transition density. The other half comes from
 the contributions of quasiparticle states with
 higher orbital angular
 momenta $l> 5$, which are all unbound continuum states.
 A remarkable feature is that at large distances, $r\gesim 8$ fm, 
 contributions from different $l$'s are similar in magnitude,
 depending only weakly on $l$, and their contributions are 
 accumulated coherently to build up the total transition
 density of the pair-addition vibration.

 The coherent contribution up to high orbital angular momenta
 is a signature of the di-neutron correlation. Let us recall that 
 a two-particle wave function made of the 
 $J=0$-coupled single-particle states brings about 
 an angular correlation
 $\sum_{m}Y^{*}_{lm}(\hat{\vecr}_{1})Y_{lm}(\hat {\vecr}_{2})
 \sim P_{l}({\rm cos}\theta_{12})$ with respect to the 
 relative angle $\theta_{12}$ between
 the positions $\hat{\vecr}_{1}$ and $\hat {\vecr}_{2}$ of the
 two particles. Since $P_{l}({\rm cos}\theta_{12})$ is peaked at
 $\theta_{12}=0$ and always positive
 for $\theta_{12} \lesim 1/l$, and 
 if we superpose them coherently over a large number
 of $l$ in a range $0\le l \lesim l_{\rm corr}$, the obtained
 two-particle wave function may exhibit an angular correlation at
 small relative angles $\theta_{12} \lesim 1/l_{\rm corr}$. 
 The validity of this argument is 
 confirmed for the Cooper pair wave function and
 the pair density $\tilde{\rho}(r)$ 
 in the HFB ground state\cite{Matsuo05,Matsuo12PhysScr}. Extending
 this argument to the pair-addition transition density
 $P^{({\rm ad})}_{L\nu}(r)$, we deduce that the 
 coherent contribution up to high orbital angular momenta
 suggests the di-neutron correlation, a spatial correlation at short 
 distances in the pair-addition transition density\cite{Serizawa09}.
 In other words,
 it suggests transfer of ``a di-neutron'' in exciting the 
 low-lying pair vibrational mode in $^{134}$Sn.

 Concerning the pair vibration in $^{120}$Sn, Fig. \ref{fig:TrD-lcutpv}(b),
 we see some coherent contributions of high-$l$ quasiparticles
 with $5< l \lesim 8-9$, but it is to much lesser extent than that in
 $^{134}$Sn. The difference is very clear for the tail region 
 $r\gesim 8$ fm;
 in $^{134}$Sn the coherent high-$l$ contribution is evident 
 while in $^{120}$Sn the transition density itself is vanishing.

\section{Giant pair vibrations in $A<132$ isotopes \label{sec:gpv}}

\begin{table}[tp]
\begin{center}
\begin{tabular}{cccrr}
 \hline\hline
&&$E_{i}+E_{i'}$&$X^{\nu}_{ii'}$&$Y^{\nu}_{ii'}$ \\
 \hline
$^{120}$Sn
&$[2f_{7/2}]^{2} $   &13.53 & 0.936 & 0.0005 \\
&$[1h_{9/2}]^{2} $   &19.77 &-0.151 &-0.0006 \\
$\hbar\omega_{\nu}$=12.36 MeV
&$[1i_{13/2}]^{2}$   &22.41 & 0.145 & 0.0007 \\
GPV1
&$[2f_{5/2}]^{2} $   &18.61 &-0.122 &-0.0003 \\
&$[3p_{3/2}]^{2} $   &16.27 & 0.116 & 0.0002 \\
&$[1h11_{3/2}]^{2} $  & 3.60 &-0.114 & 0.0129 \\
 \cline{2-5}
&$[3p_{3/2}]^{2}$    &16.27 & 0.877 & 0.0003 \\
$\hbar\omega_{\nu}$=15.88 MeV
&$[2f_{7/2}]^{2}$    &13.53 &-0.228 & 0.0015 \\
GPV2
&$[2d_{5/2}][4d_{5/2}]$ &16.29 & 0.165 & 0.0015 \\
&$[3p_{1/2}]^{2}$    &17.14 &-0.156 &-0.0001 \\
&$[1h_{9/2}]^{2}$    &19.77 &-0.154 &-0.0019 \\
&$[2f_{5/2}]^{2}$    &18.61 &-0.154 &-0.0006 \\
&$[1i_{13/2}]^{2}$   &22.41 & 0.121 &-0.0023 \\
 \hline\hline
\end{tabular}
  \caption{ \label{tabl:sn120gpvXY}
 The forward amplitudes $X_{ii'}^{\nu}$ and the backward amplitudes
 $Y_{ii'}^{\nu}$
 for the giant pair vibrations, GPV1 ($\hbar\omega_{\nu}= 12.36$
 MeV) and GPV2 ($\hbar\omega_{\nu}=15.88$ MeV), in $^{120}$Sn.
 The largest two-quasiparticle components satisfying 
 $|X_{ii'}^{\nu}|>0.1$ are listed. 
}
\end{center}
\end{table}

 Let us analyze the high-lying pair-addition modes
 GPV2 at $\hbar\omega_{\nu}=15.88$ MeV and
 GPV1 at $\hbar\omega_{\nu}=12.36$ MeV 
 in $^{120}$Sn (cf. Fig. \ref{fig:SF-gpv}(b)). 

 We first note that the pair-addition strengths of these modes,
 $B({\rm Pad}0)=1.374$ of GPV2 and 
 $B({\rm Pad}0)=2.332$ of GPV1, are
 about eight times larger than the single-particle values
 $B_{s.p.}({\rm Pad}0)=0.16$ for $j=3/2$ and 
 $B_{s.p.}({\rm Pad}0)=0.32$ for $j=7/2$, respectively.

 Let us first examine the phonon amplitudes.
 Table \ref{tabl:sn120gpvXY} shows the forward amplitudes
 $X_{ii'}^{\nu}$ of the dominant two-quasiparticle components
 of GPV1 and GPV2. Here the two-quasiparticle configurations
 with large amplitudes $|X_{ii'}^{\nu}|>0.1$ are plotted. 
 All these dominant components,
 except $[2d_{5/2}][4d_{5/2}]$, 
 are made of the quasiparticle states 
 $[2f_{7/2}]$, $[3p_{3/2}]$, 
 $[3p_{1/2}]$, $[1h_{9/2}]$, $[2f_{5/2}]$ and $[1i_{13/2}]$,
 which all belong to the shell next to the valence shell,
 i.e. the one above the $N=82$ shell gap (cf. Fig. \ref{fig:spe}).
 This confirms that GPV1 and GPV2 are indeed the giant pair vibrations.
 (Note that the two-quasiparticle configuration $[2d_{5/2}][4d_{5/2}]$
 is a particle-hole excitation from the bound $2d_{5/2}$ orbit to
 a discretized continuum state in the partial wave $d_{5/2}$. 
 This component contributes very little to the pair-transfer mode.)
 Even though the largest components $[2f_{7/2}]^{2}$ in GPV1 and
 $[3p_{3/2}]^{2}$ in GPV2 have predominant amplitudes
 $X_{ii'}^{\nu}=$ 0.94 and 0.88, respectively,
 the number ($\sim$ seven) of large components which satisfy 
 $|X_{ii'}^{\nu}|>0.1$ show the degree of collectivity 
 similar to that of the pair vibration in $^{134}$Sn.

\begin{figure}[tp]
\includegraphics[width=7.9cm]{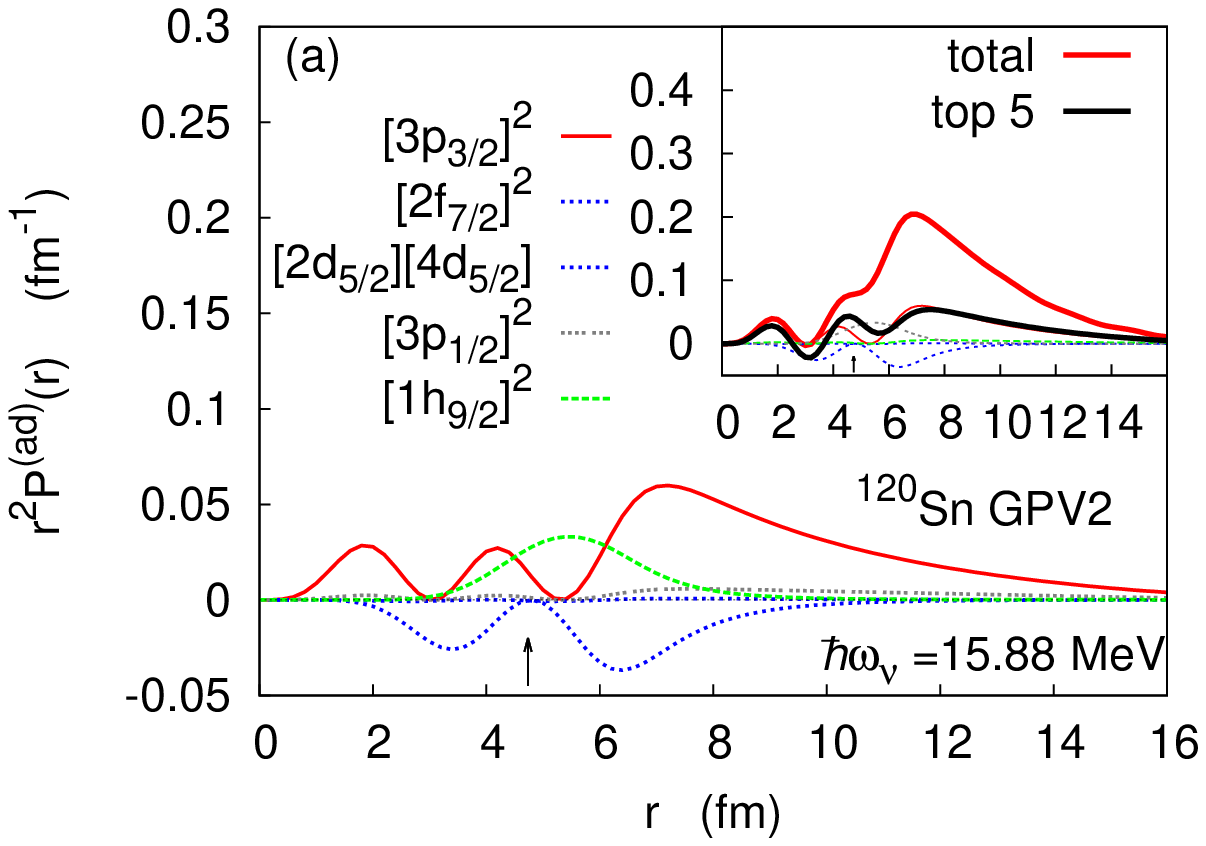}
\includegraphics[width=7.9cm]{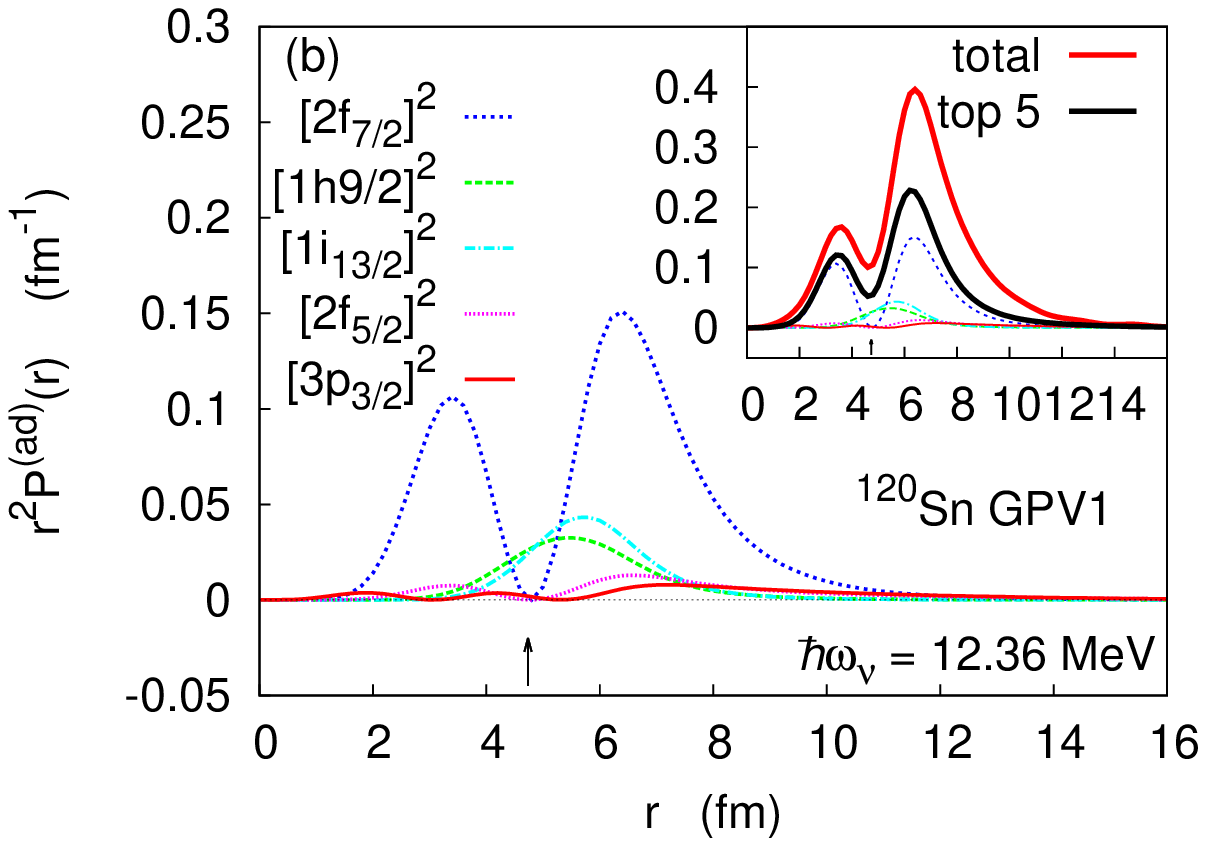}
\caption{\label{fig:TrD-Sn120top}(Color online)
 (a) The decomposed transition densities 
 $r^{2}P^{({\rm ad})}_{\nu L,ii'}(r)$ 
 for the largest two-quasiparticle components of the 
 giant pair vibration GPV2
 at $\hbar\omega_{\nu}=15.88$ MeV in $^{120}$Sn
 (cf. Table \ref{tabl:sn120gpvXY}).
 In the inset, shown also are
 the total transition density
 $r^{2}P^{({\rm ad})}_{\nu L}(r)$ and a partial sum
 of the decomposed transition densities of the largest 
 five two-quasiparticle components. 
 The arrow indicates the neutron rms radius
 $R_{N,rms}=4.73$ fm.
 (b) The same as (a) but for the giant pair vibration GPV1
 at $\hbar\omega_{\nu}=12.36$ MeV in $^{120}$Sn.
}
\end{figure}

 Figures \ref{fig:TrD-Sn120top}(a) and \ref{fig:TrD-Sn120top}(b)
 show the transition densities 
 $P^{({\rm ad})}_{\nu L}(r)$ of the GPV2 and GPV1 modes,
 respectively. It is seen that the transition densities
 extend to far outside the nuclear surface, reaching
 $r\sim 16$ and $r \sim 14$ fm, respectively.
 Plotted are also the
 decomposed transition densities
 $P^{({\rm ad})}_{\nu L,ii'}(r)$ for the largest five
 components listed in Table \ref{tabl:sn120gpvXY}.
 In the GPV2 mode (see the inset of
 Fig. \ref{fig:TrD-Sn120top}(a)), the amplitude of 
 transition density of the main component $[3p_{3/2}]^{2}$ is
 significantly smaller than that of the 
 total transition density. Even if we consider
 the superposition of the decomposed transition densities with the largest
 five components, it reproduces only one-quarter of the total
 transition density of the GPV2 mode. A similar situation is observed
 also in the case of GPV1 (Fig. \ref{fig:TrD-Sn120top}(b)), where
 the most dominant component $[2f_{7/2}]^{2}$ accounts for 
 only one-third of the maximum of the total transition density,
 and a sum of the largest five components give only a bit more than
 a half.
 The large components alone are not sufficient to account for the
 collectivity of GPV1 and GPV2, especially that of GPV2.

\begin{figure}[tp] 
\includegraphics[width=7.9cm]{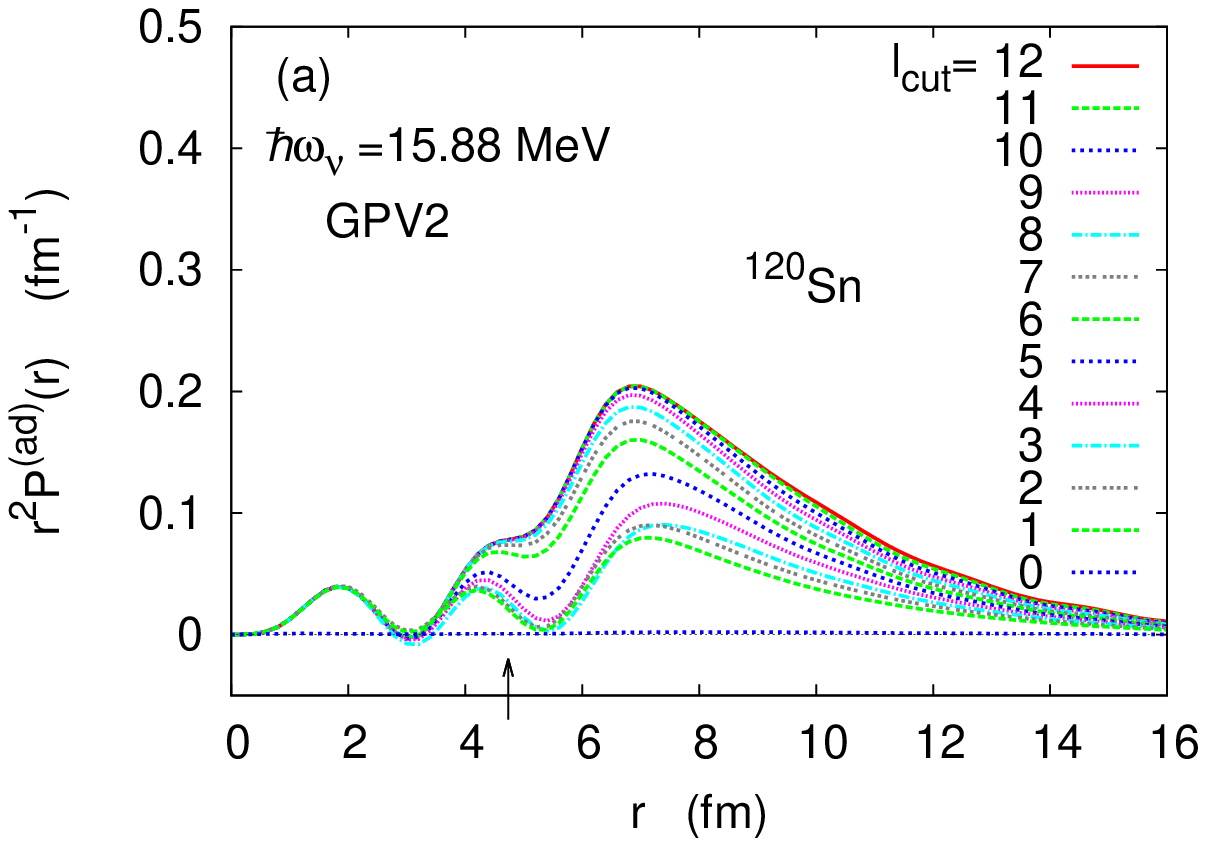}
\includegraphics[width=7.9cm]{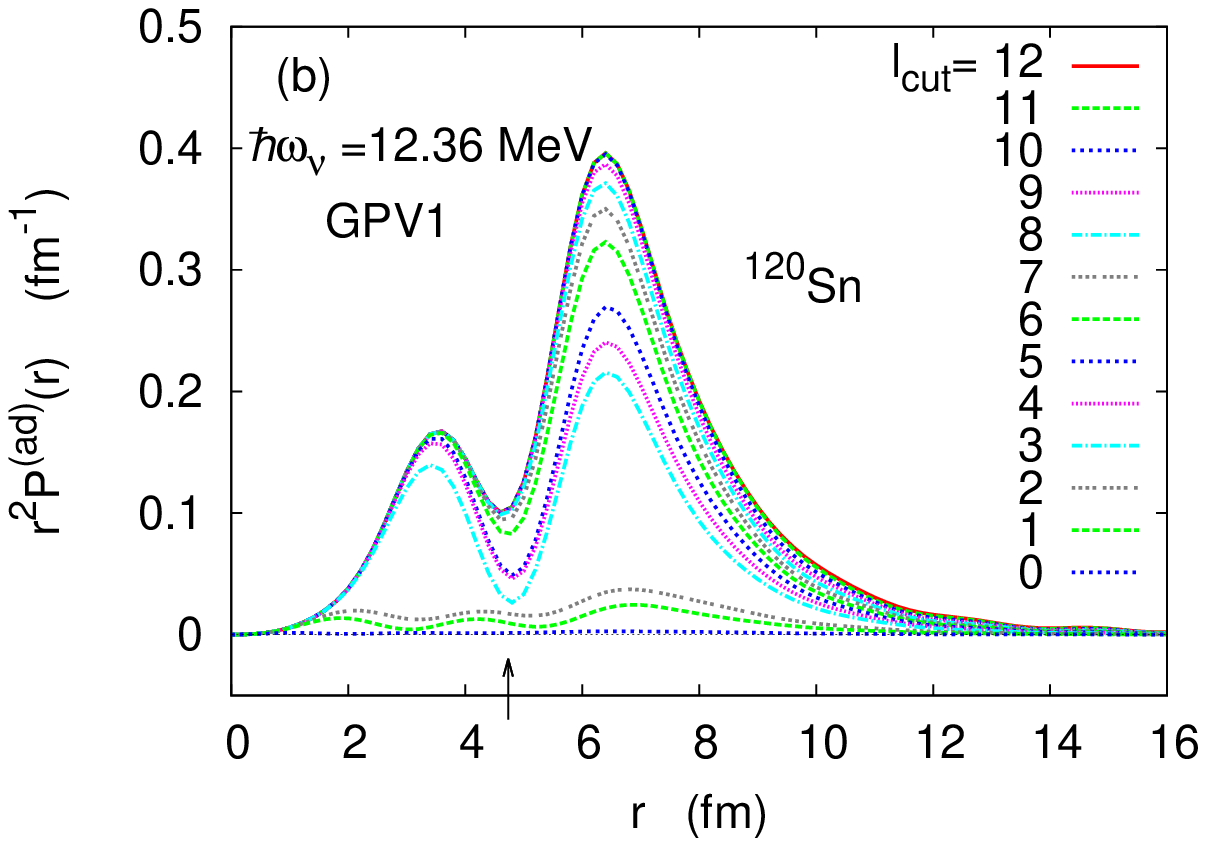}
\caption{\label{fig:TrDl-gpv}(Color online)
 The $l$-cutted transition densities $r^{2}P^{({\rm ad})}_{\nu L,l_{cut}}(r)$
 with $l_{cut}$=0, 1, 2, $\cdots$,12 
 for the giant pair vibrations
 (a) GPV2 ($\hbar\omega_{\nu} = 15.88$ MeV) and (b) GPV1
 ($\hbar\omega_{\nu} =12.36$ MeV) in $^{120}$Sn.
 The arrow indicates the neutron rms radius $R_{N,rms}$.
}
\end{figure}

 We also analyze the microscopic structure of these modes
 in terms of the orbital angular momentum $l$ of the
 quasiparticle orbits. The method is the same as what
 we have done for the low-lying pair vibration, i.e. 
 we plot partial sums of
 the decomposed transition densities 
 $P^{({\rm ad})}_{\nu L,l_{cut}}(r) =
 \sum_{ii',l\le l_{cut}}P^{({\rm ad})}_{\nu L,ii'}(r)$,
 for various values of the angular momentum cut-off $l_{cut}$.
 The results are shown in Fig. \ref{fig:TrDl-gpv}.

 It is seen in both Figs. \ref{fig:TrDl-gpv} (a) and \ref{fig:TrDl-gpv}(b)
 that two-quasiparticle components with 
 different $l$'s contribute coherently, and the
 contributions from high-$l$ states (with $l>l_{occ}=5$)
 are significant. The latter feature is very strong
 in the external region $r\gesim 8$ fm of GPV2 (Fig. \ref{fig:TrDl-gpv}(a)).
 This is quite similar to the features observed for
 the low-lying pair vibration in $^{134}$Sn (Fig. \ref{fig:TrD-lcutpv}(a)).
 Applying the same arguments on Fig. \ref{fig:TrD-lcutpv}(a), we deduce that
 the di-neutron correlation appears also in the 
 GPV2 mode in the stable $^{120}$Sn isotope, and that 
 the GPV1 mode also exhibits the same feature, but to a lesser extent.
 This is in contrast to the low-lying pair vibration in $^{120}$Sn
 (Fig. \ref{fig:TrD-lcutpv}(b)),
 where the di-neutron feature is seen only weakly.

\section{Pair rotation}

\begin{figure}[tp]
\includegraphics[width=7.9cm]{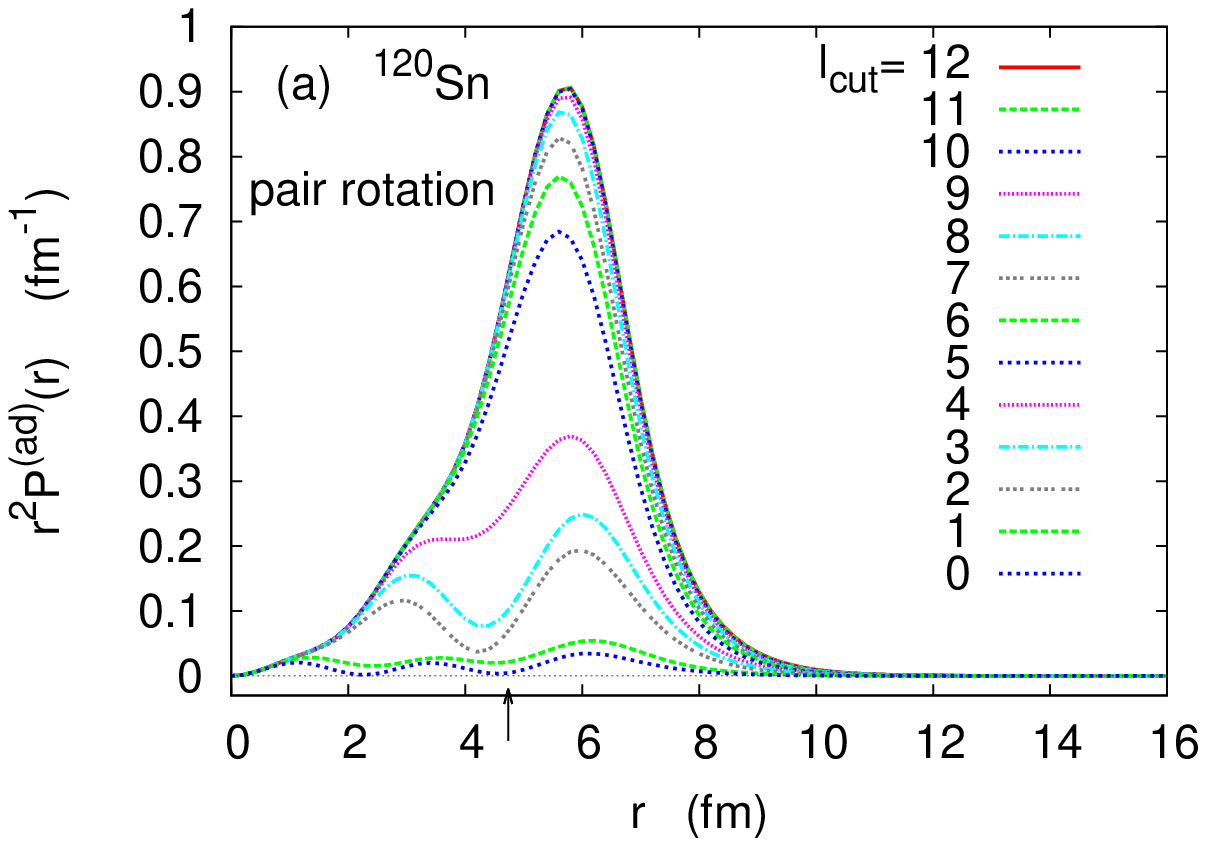}
\includegraphics[width=7.9cm]{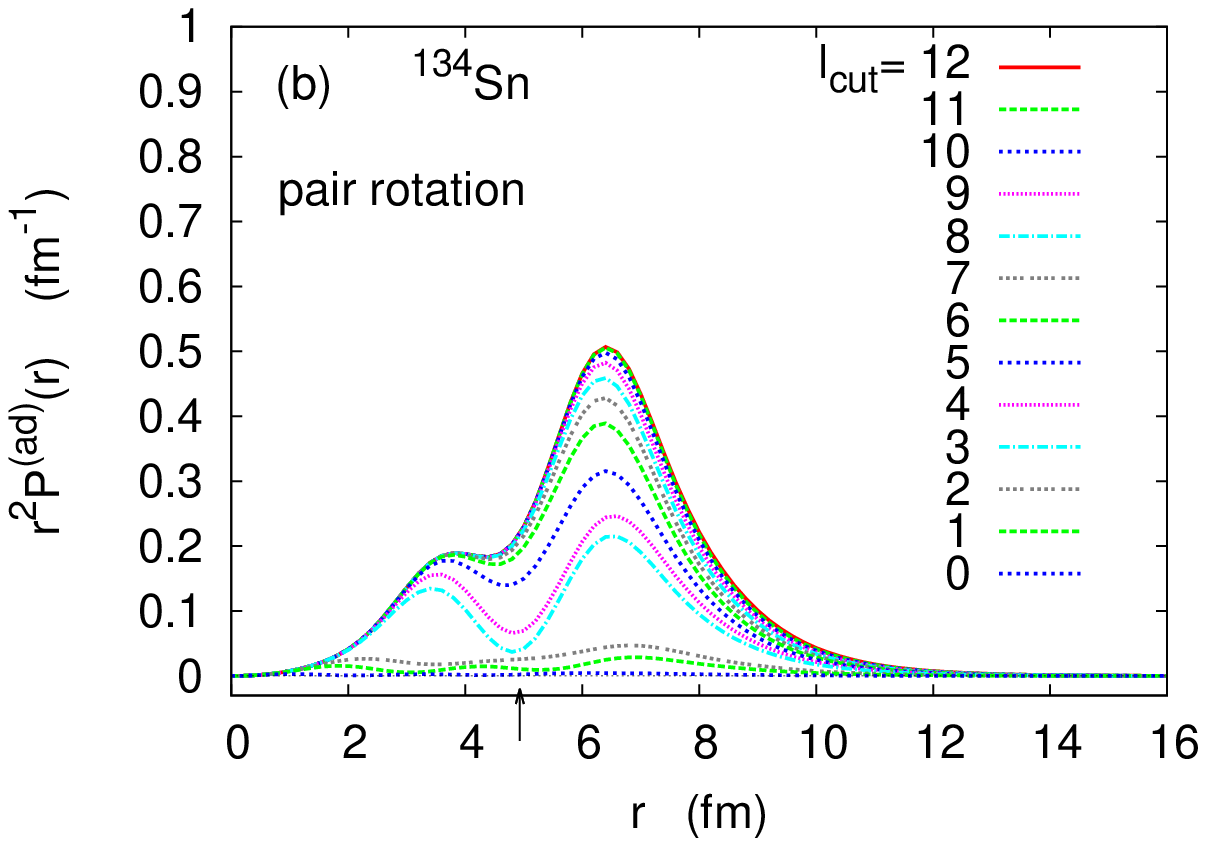}
\includegraphics[width=7.9cm]{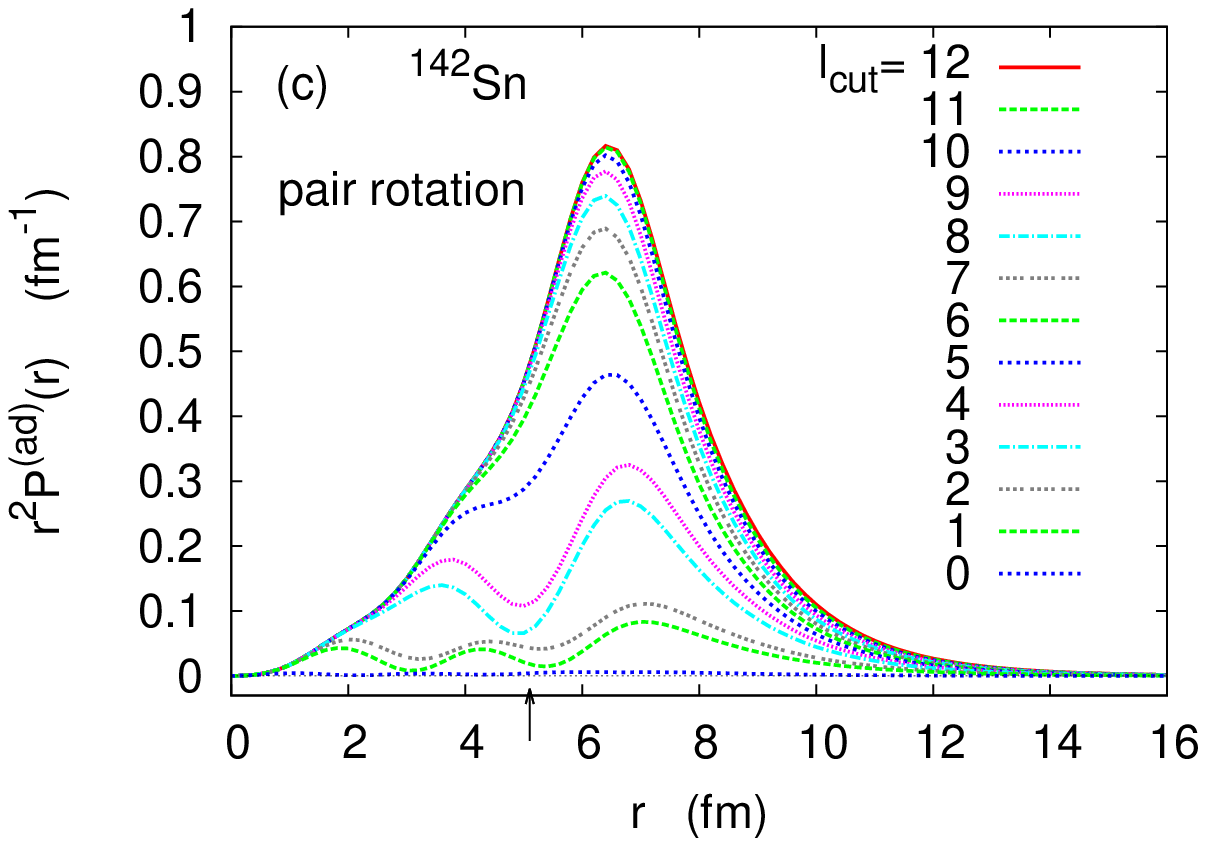}
\caption{\label{fig:TrD-lcutgs}(Color online)
 The $l$-cutted transition densities $r^{2}P^{({\rm ad})}_{{\rm gs},l_{cut}}(r)$
 with $l_{cut}$=0, 1, 2, $\cdots$,12 
 for the pair rotational modes in 
 (a) $^{120}$Sn, (b) $^{134}$Sn and (c) $^{142}$Sn.
 The arrow indicates the neutron rms radius $R_{N,rms}=4.73$ fm, $4.93$ fm
 and $5.10$ fm for $^{120}$Sn, $^{134}$Sn and $^{142}$Sn,
 respectively. 
}
\end{figure}

 In this section we show the microscopic structure
 of the pair rotation, i.e. the transfer mode populating 
 the ground state of neighboring $N+2$ isotope.

 We decompose
 the transition density $P_{{\rm gs}}^{({\rm ad})}(r)$
 with respect to the quasiparticle states involved, 
 as is shown in Eqs.(\ref{eq:TrDgs4}) and (\ref{eq:TrDgs5}). We then
 introduce a partial
 sum specified by the orbital angular momentum cutoff $l_{cut}$ as
 $P_{{\rm gs},l_{cut}}^{({\rm ad})}(r) =\sum_{i,l<l_{cut}}
 P_{{\rm gs},ii}^{({\rm ad})}(r)$ (cf Eq.(\ref{eq:TrDgs5})).
 This decomposition is shown in 
 Fig. \ref{fig:TrD-lcutgs}(a)-(c)
 for the ground-state transfer of $^{120}$Sn, $^{134}$Sn and $^{142}$Sn,
 respectively.

 The amplitude in the external region develops significantly with
 increasing the neutron number, from $^{120}$Sn to $^{134}$Sn and 
 $^{142}$Sn. In addition to this overall
 feature (discussed already in Ref.\cite{Shimoyama11}), we find here
 that the high-$l$ contribution (with $l\ge 5$) to the
 transition density grows with increasing the neutron number. 
 Looking at $r=6$ fm where the plotted transition density is close to the
 largest, the high-$l$ contribution is $\sim 20\%, \sim 30\%$ and $\sim 40\%$
 in $^{120}$Sn, $^{134}$Sn and $^{142}$Sn, respectively.
 It is also obvious that all high-$l$'s contribute coherently
 to build up the pair-addition transition density 
 $P_{{\rm gs}}^{({\rm ad})}(r)$. 
 In short, the features suggesting the di-neutron correlation are 
 seen well in the pair rotational mode in 
 $^{134}$Sn, and especially in $^{142}$Sn, but not very
 significantly in $^{120}$Sn.

 It is noted here that the Fermi energies in 
 $^{134}$Sn and $^{142}$Sn are small ($\lambda_n=-2.56$
 and $-1.42$ MeV, respectively), and the lowest energy
 quasiparticle states in these isotopes are $3p_{1/2}, 3p_{3/2}$
 and $2f_{7/2}$ states (cf Fig. \ref{fig:spe}). However, 
 the contribution of these quasiparticle
 states to the pair-addition transition density is not very large as
 seen in the curve with $l_{cut}=3$. 
 The high-$l$ quasiparticle states, which give dominant
 part of the transition density in the external region, are
 all unbound quasiparticle states in the continuum energy region. 
 This is one of the mechanisms that make the transition density
 very extended in $^{134}$Sn and $^{142}$Sn. 

 All the above mentioned
 features of the pair rotation
 in $^{134}$Sn and $^{142}$Sn 
 have some similarity to those of the low-lying pair vibration in
 $^{134}$Sn and the GPV's in the isotopes with $A<132$, discussed
 in the previous sections.

\section{systematics}

 In $^{110-130}$Sn, we 
 systematically observe two peaks of the high-lying giant pair vibrations,
 GPV1 and GPV2. 
 In order to clarify the isotopic trends of 
 the low-lying pair vibration, the
 giant pair vibration and the pair rotation, we have 
 performed systematic numerical calculations
 for all the even-$N$
 isotopes from $A=120$ to $A=150$. 
 In $^{132}$Sn where the $N=82$ shell is closed and
 the neutron pair gap vanishes, we find two low-lying pair vibration
 in the pair-addition channel. The lowest one, denoted PV1 hereafter,
 can be regarded as 
 a pair-addition mode populating the ground state of $^{134}$Sn,
 while the second one, denoted PV2, is 
 another pair-addition mode populating the second
 $0^+$ state in $^{134}$Sn. In $^{134-140}$Sn we find
 the characteristic low-lying pair-addition vibration which we discussed
 in section IV and in Ref.\cite{Shimoyama11}. In $^{142-150}$Sn,
 the pairing collective mode having a large transfer strength is only
 the pair rotation.
 (There exists low-lying monopole modes having very small strength 
 in $^{142-150}$Sn
 (cf Fig. \ref{fig:st-Sn}), but we do not discuss these in this paper.)
 
 From the systematic analysis, we find the following series and
 relations connecting the above mentioned various pair-transfer modes
 that appear in different regions of the Sn isotopes. 

\begin{enumerate}

\item[i)] The giant pair vibration GPV2 in $^{120-130}$Sn,
 the second excited pair vibration PV2 in $^{132}$Sn, 
 and the low-lying pair vibration in $^{134-140}$Sn.

\item[ii)] The giant pair vibration GPV1 in $^{120-130}$Sn,
 the lowest pair vibration PV1 in $^{132}$Sn, 
 and the pair rotation in $^{134-140}$Sn.

\item[iii)] The pair rotation in $^{142-150}$Sn and the
 above two series i) and ii).

\end{enumerate}

\subsection{Series i)}

\begin{table*}[t]
\begin{center}
\begin{tabular}{cc|rrrrrr|r|rrrr}
 \hline \hline
&            & \multicolumn{6}{c}{GPV2} &\multicolumn{1}{|c|}{PV2} &\multicolumn{4}{c}{low-lying pair vibration} \\
 \hline
&            &$^{120}$Sn&$^{122}$Sn&$^{124}$Sn&$^{126}$Sn&$^{128}$Sn&$^{130}$Sn&$^{132}$Sn&$^{134}$Sn&$^{136}$Sn&$^{138}$Sn&$^{140}$Sn\\
&$\hbar\omega_{\nu}$  (MeV)
            & 15.88  & 15.12  & 14.42  & 13.81  & 13.27  & 12.78  & 8.35   & 3.81   & 3.21   & 2.46  & 1.38   \\ 
&$B$(Pad0)  (fm$^{0}$) & 1.374  & 1.545  & 1.682  & 1.816  & 2.023  & 2.248  & 2.928  & 3.233  & 3.721  & 4.615  & 2.932  \\ 
 \hline
&$[3p_{3/2}]^{2}$    & 0.877  & 0.896  & 0.886  & 0.870  & 0.855  & 0.830  & 0.768  & 0.752  & 0.723  & 0.676  & 0.479  \\ 
&$[2f_{7/2}]^{2}$    & -0.228  & -0.228  & -0.230  & -0.236  & -0.245  & -0.259  & -0.296  & -0.297  & -0.302  & -0.323  & -0.764  \\ 
&$[2d_{5/2}][4d_{5/2}]$ & 0.165  &     &     &     &     &     &     &     &     &     &     \\ 
&$[3p_{1/2}]^{2}$    & -0.156  & -0.158  & -0.159  & -0.161  & -0.164  & -0.167  & -0.172  & -0.177  & -0.185  & -0.193  & -0.149  \\ 
&$[1h_{9/2}]^{2}$    & -0.154  & -0.174  & -0.193  & -0.214  & -0.243  & -0.282  & -0.344  & -0.368  & -0.402  & -0.449  & -0.360  \\ 
&$[2f_{5/2}]^{2}$    & -0.154  & -0.169  & -0.183  & -0.198  & -0.217  & -0.240  & -0.276  & -0.279  & -0.289  & -0.309  & -0.239  \\ 
&$[1i_{13/2}]^{2}$   & 0.120  & 0.131  & 0.142  & 0.154  & 0.171  & 0.193  & 0.231  & 0.241  & 0.259  & 0.288  & 0.229  \\ 
$X_{ii'}^{\nu}$
&$[3p_{3/2}][4p_{3/2}]$ & 0.088  & 0.092  & 0.093  & 0.094  & 0.095  & 0.098  & 0.103  & 0.101  & 0.100  & 0.100  & 0.071  \\ 
&$[1g_{7/2}][3g_{7/2}]$ & -0.086  &     &     &     &     &     &     &     &     &     &     \\ 
&$[3p_{1/2}][4p_{1/2}]$ & -0.053  & -0.056  & -0.058  & -0.059  & -0.060  & -0.062  & -0.067  & -0.066  & -0.067  & -0.069  & -0.051  \\ 
&$[2f_{5/2}][3f_{5/2}]$ &     & -0.045  & -0.047  & -0.044  & -0.049  & -0.051  & -0.056  & -0.056  & -0.058  & -0.061  & -0.045  \\ 
&$[1h_{11/2}]^{2}$   &     & -0.045  &     &     &     &     &     &     &     &     &     \\ 
&$[2d_{3/2}][4d_{3/2}]$ &     &     & 0.071  & -0.093  &     &     &     &     &     &     &     \\ 
&$[2d_{5/2}][3d_{5/2}]$ &     &     &     &     & 0.054  &     &     &     &     &     &     \\ 
&$[2f_{7/2}][3f_{7/2}]$ &     &     &     &     &     & 0.053  & 0.060  &     &     &     &     \\ 
&$[3g_{9/2}]^{2}$    &     &     &     &     &     &     &     & 0.047  & 0.051  &     & 0.045  \\ 
&$[2j_{15/2}]^{2}$   &     &     &     &     &     &     &     &     &     & 0.058  &     \\ 

 \hline \hline
\end{tabular}
 \caption{ \label{tabl:Sngpv2X}
 The forward amplitudes $X_{ii'}^{\nu}$ of
 the giant pair vibration GPV2
 in $^{120-130}$Sn, of the second excited pair vibration PV2 in $^{132}$Sn,
 and of the low-lying pair vibration in $^{134-140}$Sn.
 The two-quasiparticle components having ten largest values of
 $|X_{ii'}^{\nu}|$ are listed for each mode. 
 The excitation energy $\hbar\omega_{\nu}$ and the strength 
 $B({\rm Pad0};{\rm gs}\rightarrow \nu)$ are also listed.
}
\end{center}
\end{table*}

 We first discuss the series i). 
 The forward amplitudes $X^{\nu}_{ii'}$ of the GPV2 in $^{120-130}$Sn,
 the second excited pair vibration PV2 in $^{132}$Sn,
 and the low-lying pair vibration in $^{134-140}$Sn 
 are listed in Table \ref{tabl:Sngpv2X}
 for 
 the largest ten two-quasiparticle components.

 It is seen that the $X$-amplitudes of the GPV2 are very similar
 for all the isotopes in the interval $A=120-130$.
 The largest is the two-quasiparticle configuration $[3p_{3/2}]^{2}$ 
 and the other largest components are
 $[2f_{7/2}]^{2}$, $[3p_{1/2}]^{2}$, $[1h_{9/2}]^{2}$ $[2f_{5/2}]^{2}$
 and $[1i_{13/2}]^{2}$.
 The $X$-amplitudes of them all vary only gradually
 with changing the neutron number. 
 A slight increase of
 collectivity is seen; for instance, the
 $X$-amplitude of $[1i_{13/2}]^{2}$, the fifth largest in most cases,
 increases from $0.120$ in $^{120}$Sn to $0.193$ in $^{130}$Sn. 
 This is consistent
 with the increase of the pair-addition strength $B({\rm Pad}0)$
(cf . Fig. \ref{fig:st-Sn} and Table \ref{tabl:Sngpv2X}).

 We now compare the $X$-amplitudes of the GPV2 in $^{128,130}$Sn with
 those of the second pair vibration PV2
 in $^{132}$Sn and the low-lying pair vibration in
 $^{134}$Sn. A similarity
 is obvious for the $X$-amplitudes of the largest components
 $[3p_{3/2}]^{2}$, $[2f_{7/2}]^{2}$,
 $[3p_{1/2}]^{2}$, $[1h_{9/2}]^{2}$, $[2f_{5/2}]^{2}$ and
 $[1i_{13/2}]^{2}$; 
 the $X$-amplitudes are smoothly connected among
 the GPV2 in $^{120-130}$Sn,
 PV2 in $^{132}$Sn and the low-lying pair vibration in $^{134}$Sn.
 Further increasing the neutron number to
 $A=136, 138$, we
 find a continuation of the similarity for the
 low-lying pair-vibration in $^{136,138}$Sn. 
 In $A=140$, we can still trace the same similarity to some extent, 
 but at the same time the variation from $A=138$ to $A=140$ becomes
 slightly large.

\begin{figure}[tp] 
\includegraphics[width=7.9cm]{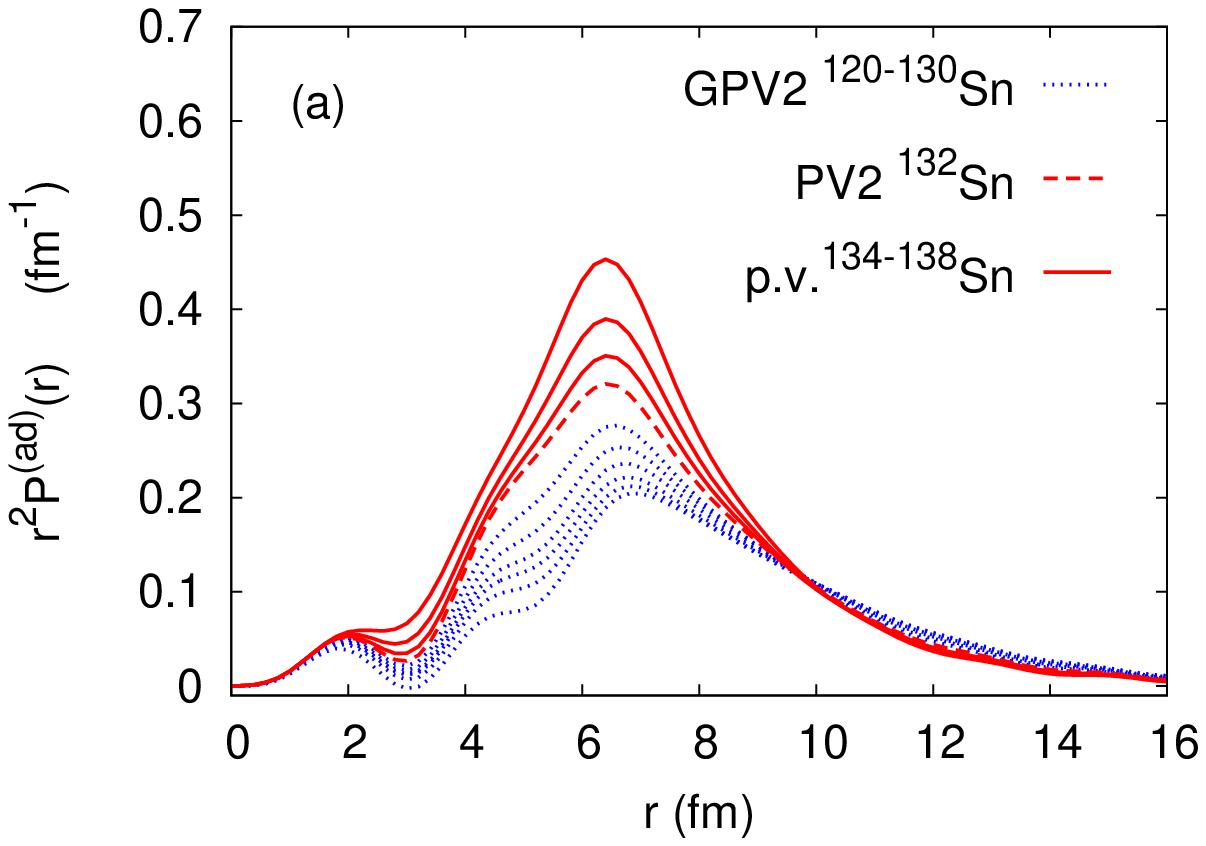}
\includegraphics[width=7.9cm]{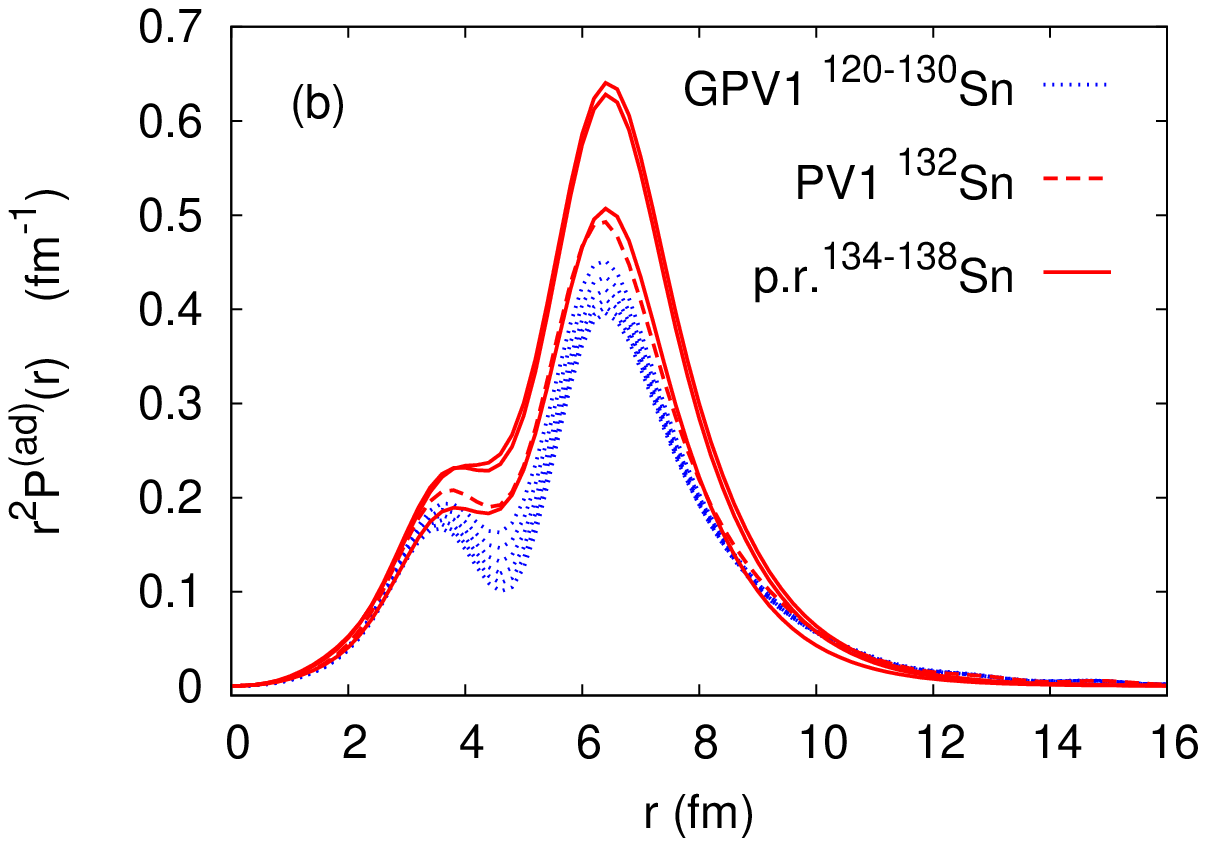}
\caption{\label{fig:TRD-sn-gspvgpv}(Color online)
 (a) The transition densities $r^{2}P_{\nu L}^{({\rm ad})}(r)$ for the
 giant pair vibration GPV2 in $A=$120-130 (dotted),
 for the second pair vibration PV2 in $A=132$ (dashed) and  
 for the low-lying pair vibration in $A=134-138$
 (solid).
 (b) The transition densities for the giant pair vibration GPV1 in 
 $A=120-130$ (dotted), for the first pair vibration
 PV1 in $A=132$ (dashed) and the transition densities of the pair rotation
 $r^{2}P_{\rm gs}^{({\rm ad})}(r)$ in $A=134-138$
 (solid).
}
\end{figure}

 Figure \ref{fig:TRD-sn-gspvgpv}(a) shows 
 the pair-addition transition densities for the series.
 We can confirm that the transition densities
 of the GPV2 modes in $^{120-130}$Sn are smoothly
 connected to that for the pair-addition vibration
 in $^{132-138}$Sn.

\subsection{Series ii)}

\begin{table*}[t]
\begin{center}
\begin{tabular}{cc|rrrrrr|r|rrrr}
 \hline\hline
&             & \multicolumn{6}{c}{GPV1} &\multicolumn{1}{|c|}{PV1} &\multicolumn{4}{c}{pair rotation} \\
 \hline
&             &$^{120}$Sn &$^{122}$Sn &$^{124}$Sn &$^{126}$Sn &$^{128}$Sn &$^{130}$Sn &$^{132}$Sn &$^{134}$Sn &$^{136}$Sn &$^{138}$Sn &$^{140}$Sn \\
&$\hbar\omega_{\nu}$  (MeV)
             & 12.36   & 11.46   & 10.65   & 9.94   & 9.30    & 8.71   & 4.15   &\multicolumn{1}{c}{-}
                                                             &\multicolumn{1}{c}{-}
                                                                   &\multicolumn{1}{c}{-}
                                                                        &\multicolumn{1}{c}{-}\\
&$B$(Pad0)  (fm$^{0}$)  & 2.332  & 2.416   & 2.545   & 2.679   & 2.877   & 3.105  & 3.973  & 3.607   & 5.662  & 6.048  & 5.157 \\
 \hline
&$[2f_{7/2}]^{2}$     & 0.936  & 0.939  & 0.940  & 0.942  & 0.943  & 0.945  & 0.942  &      &     &     &    \\
&$[1h_{9/2}]^{2}$     & -0.151  & -0.154  & -0.157  & -0.161  & -0.167  & -0.175  & -0.192  &      &     &     &    \\
&$[1i_{13/2}]^{2}$    & 0.145  & 0.147  & 0.149  & 0.153  & 0.158  & 0.165  & 0.181  &      &     &     &    \\
&$[2f_{5/2}]^{2}$     & -0.122  & -0.123  & -0.125  & -0.128  & -0.132  & -0.136  & -0.146  &      &     &     &    \\
$X_{ii'}^{\nu}$
&$[3p_{3/2}]^{2}$     & 0.116  & 0.113  & 0.112  & 0.111  & 0.112  & 0.112  & 0.117  &      &     &     &    \\ 
&$[1h_{11/2}]^{2}$    & -0.113  & -0.106  & -0.094  & -0.077  &      &      &      &      &     &     &    \\
&$[3p_{3/2}][4p_{3/2}]$  & 0.059  & 0.057  & 0.056  & 0.054  & 0.053  & 0.053  & 0.054  &      &     &     &    \\
&$[2f_{5/2}][3f_{5/2}]$  & -0.057  & -0.055  & -0.053  & -0.051  & -0.049  & -0.048  & -0.049  &      &     &     &    \\
&$[3p_{1/2}]^{2}$     & -0.050  & -0.049  & -0.049  & -0.049  & -0.050  & -0.051  & -0.054  &      &     &     &    \\
&$[2f_{7/2}][3f_{7/2}]$  & 0.040  & 0.039  &      &      &      &      &      &      &     &     &    \\
&$[2j_{15/2}][3j_{15/2}]$ &      &      & 0.038  & 0.039  & 0.040  & 0.041  & 0.044  &      &     &     &    \\
&$[3g_{9/2}]^{2}$     &      &      &      &      & 0.038  &      &      &      &     &     &    \\
&$[2j_{15/2}]^{2}$    &      &      &      &      &      & 0.039  & 0.044  &      &     &     &    \\

 \hline \hline
\end{tabular}
 \caption{ \label{tabl:Sngpv1X}
 The forward amplitudes $X_{ii'}^{\nu}$ of
 the giant pair vibration GPV1
 in $^{120-130}$Sn and of the first pair vibration PV1 in $^{132}$Sn.
 The two-quasiparticle components having ten largest values of
 $|X_{ii'}^{\nu}|$ are listed for each mode. 
 The excitation energy $\hbar\omega_{\nu}$ and the strength 
 $B({\rm Pad0};{\rm gs}\rightarrow \nu)$ are also listed.
 For the pair rotation in $^{134-140}$Sn, we only list
 the strength $B({\rm Pad0};{\rm gs}\rightarrow {\rm gs})$ since 
 the $X$- and $Y$-amplitudes are not evaluated.
}
\end{center}
\end{table*}

 The second series is identified among
 the GPV1 modes in $^{120-130}$Sn,
 the lowest pair vibration PV1 in $^{132}$Sn, 
 and the pair rotations in the isotopes $^{134-140}$Sn. 
 The latter two populates the ground states in neighboring $N+2$ isotopes.

 We show in Table \ref{tabl:Sngpv1X} the $X$-amplitudes
 of two-quasiparticle components for the giant pair
 vibration GPV1 modes in $^{120-130}$Sn
 and for the lowest pair-addition vibration PV1
 in $^{132}$Sn.
 The main 
 components include 
 $[2f_{7/2}]^{2}$, $[1h_{9/2}]^{2}$, $[1i_{13/2}]^{2}$, 
 $[2f_{5/2}]^{2}$ and $[3p_{3/2}]^{2}$
 (they are all shared with the GPV2 mode ), and the largest is
 $[2f_{7/2}]^{2}$ (this is different from GPV2). 
 From the $X$-amplitudes of these components, we find that 
 the GPV1 modes in $^{120-130}$Sn have common microscopic
 structure with only small variation with $N$. 
 It is seen also that the lowest pair vibration mode PV1
 in $^{132}$Sn has essentially the same microscopic structure with
 that of the GPV1 modes in $^{120-130}$Sn. This mode can be regarded
 as a smooth continuation of GPV1. 

\begin{figure}[tp]
\includegraphics[width=7.9cm]{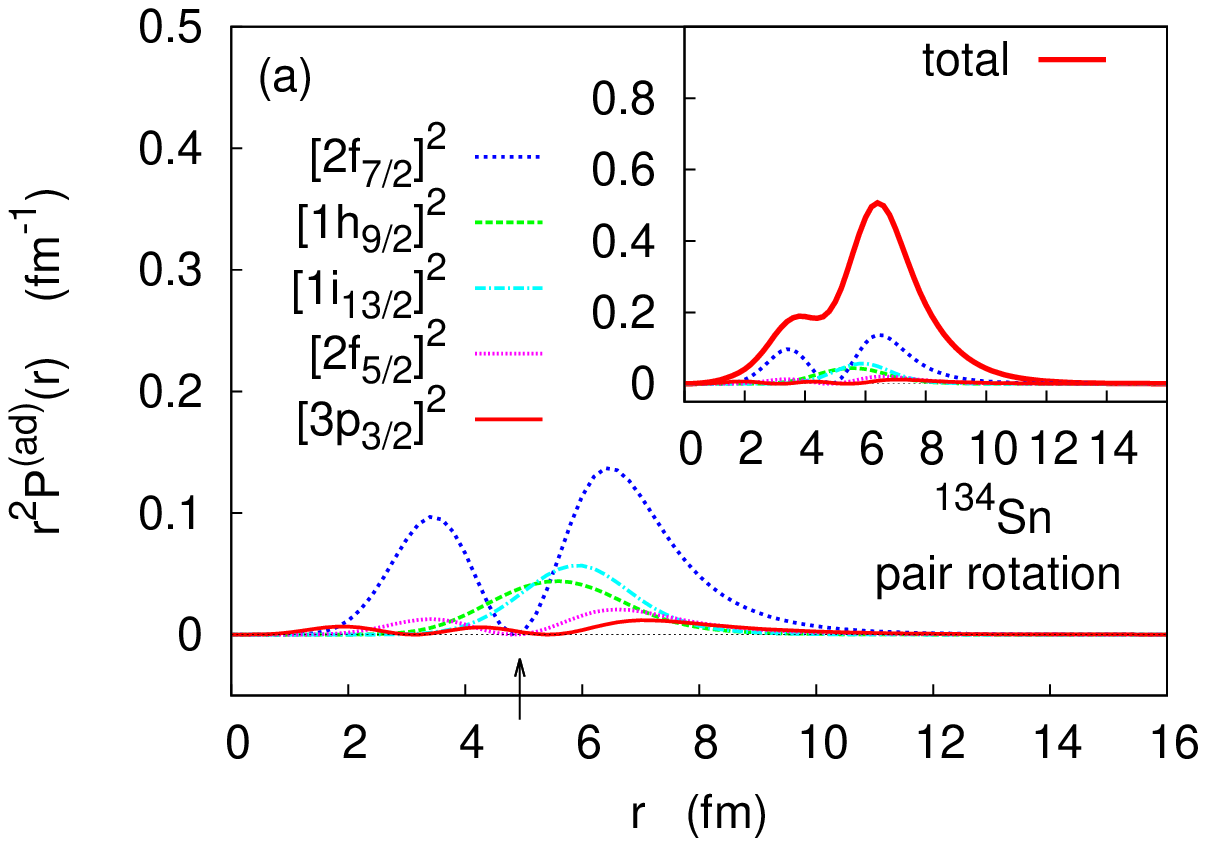}
\includegraphics[width=7.9cm]{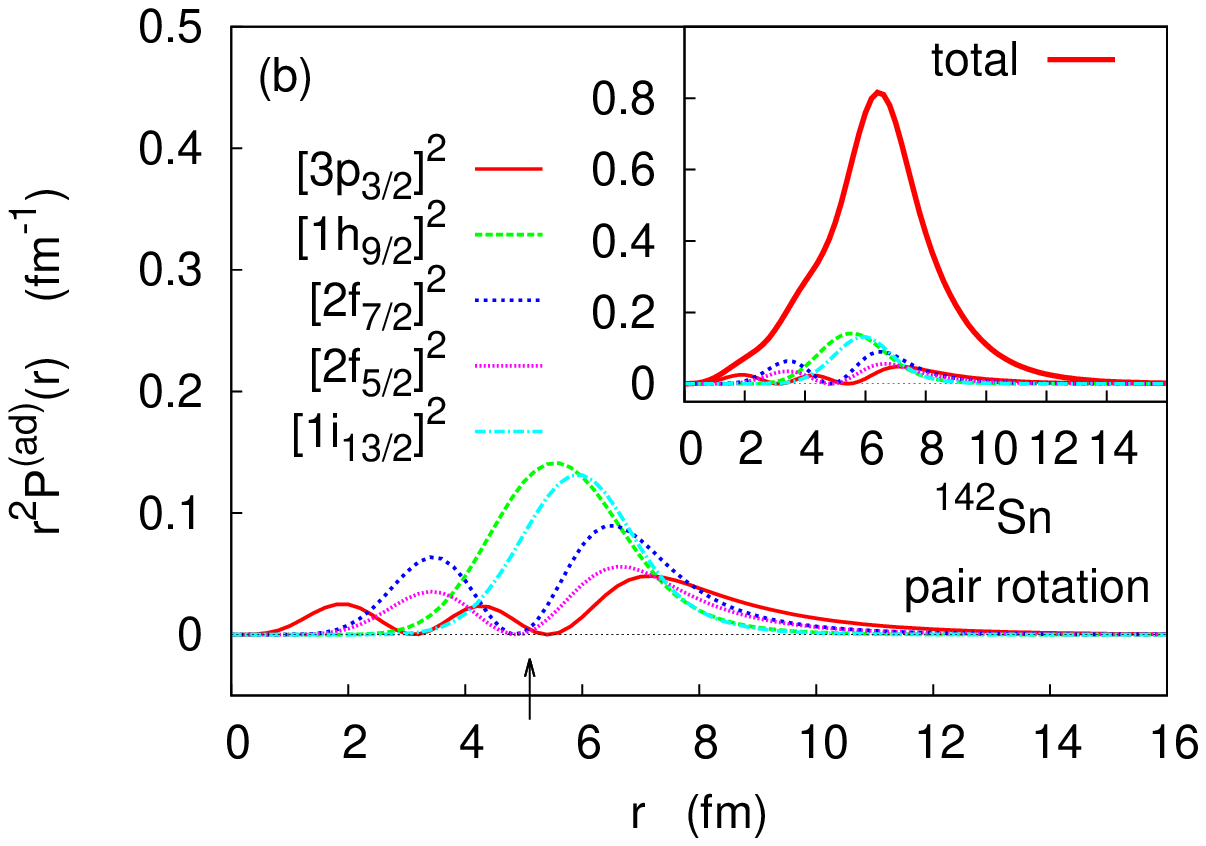}
\caption{\label{fig:TrD-gs}(Color online)
 (a) The decomposed transition densities
 $P^{({\rm ad})}_{{\rm gs},ii'}(r)$ 
 of the components 
 $[2f_{7/2}]^2$, $[1h_{9/2}]^2$, $[1i_{13/2}]^2$, $[2f_{5/2}]^2$,
 and $[3p_{3/2}]^2$ for the pair rotation in $^{134}$Sn.
 These five components correspond to those plotted in 
 Fig. \ref{fig:TrD-Sn120top}(b).
 The inset also shows the total transition density
 $P^{({\rm ad})}_{\rm gs}(r)$.
 The arrow indicates the neutron rms radius $R_{N,rms}$.
 (b) The same as (a) but for the pair rotation 
 in $^{142}$Sn. The components plotted are
 $[3p_{3/2}]^{2}$, $[1h_{9/2}]^{2}$, $[2f_{7/2}]^{2}$, $[2f_{5/2}]^{2}$
 and $[1i_{13/2}]^{2}$.
}
\end{figure}

 For the pair rotation modes in $^{134-140}$Sn, the $X$- and $Y$-amplitudes
 are not evaluated. In place of them, we calculate
 decomposed transition densities 
 $P_{{\rm gs},ii}^{({\rm ad})}(r)$ (Eq.(\ref{eq:TrDgs5})) for each of
 the main configurations 
 $[2f_{7/2}]^{2}$, $[1h_{9/2}]^{2}$, $[1i_{132}]^{2}$,
 $[2f_{5/2}]^{2}$ and $[3p_{3/2}]^{2}$ of GPV1 in $^{120-130}$Sn. 
 They are shown in Fig. \ref{fig:TrD-gs}(a). Comparing with the
 decomposed transition densities of GPV1, plotted in
 Fig. \ref{fig:TrD-Sn120top}(b) for $^{120}$Sn, we find a close
 similarity between the two figures. We find also that the pair
 rotations in $^{134-140}$Sn are all similar (the decomposed transition
 densities for $^{136-140}$Sn are not shown here). 
 It is seen in Fig. \ref{fig:TRD-sn-gspvgpv}(b) that the transition densities
 of the GPV1 modes in $^{120-130}$Sn smoothly
 match that of the PV1 in $^{132}$Sn, and those of the
 pair rotations in $^{134-138}$Sn.

\subsection{Relation iii)}

 Finally we discuss the relation iii). As a representative
 of the pair rotational modes in $^{142-150}$Sn,we take $^{142}$Sn and show
 in Fig. \ref{fig:TrD-gs}(b) microscopic decomposition of 
 the pair-addition transition density $P_{{\rm gs},ii}^{({\rm ad})}(r)$
 for the quasiparticle configurations $[3p_{3/2}]^{2}$, $[1h_{9/2}]^{2}$,
 $[2f_{7/2}]^{2}$, $[2f_{5/2}]^{2}$ and $[1i_{13/2}]^{2}$.
 We now compare this with the decomposed transition densities
 of the low-lying pair vibration and the pair rotation modes in $^{134}$Sn
 (Fig. \ref{fig:TrD-Sn120134top}(a) and Fig. \ref{fig:TrD-gs}(a)).
 It is not difficult to find some similarity between
 the transition density of the pair rotation in $^{142}$Sn
 (Fig. \ref{fig:TrD-gs}(b))
 and that of the low-lying pair vibration in $^{134}$Sn
 (Fig. \ref{fig:TrD-Sn120134top}(a)). To be more specific, 
 the components which have the largest amplitudes at $r\sim 6$ fm 
 are the quasiparticle
 states $[1h_{9/2}]$ and $[1i_{13/2}]$ in both cases. 
 The largest contribution in the external region $r\gesim 8$ fm
 is that of the quasiparticle state $[3p_{3/2}]$.
 However, a difference is seen in the contribution of $[2f_{7/2}]$, which
 has positive amplitude for the pair rotation in $^{142}$Sn
 (Fig. \ref{fig:TrD-gs}(b)),
 but the phase is opposite for the low-lying pair vibration in
 $^{134}$Sn (Fig. \ref{fig:TrD-Sn120134top}(a)).
 We then compare the pair rotation in $^{134}$Sn
 (Fig. \ref{fig:TrD-gs}(a)) and that in $^{142}$Sn
 (Fig. \ref{fig:TrD-gs}(b)), and find 
 that the $[2f_{7/2}]$ amplitude is positive and large in both cases. The 
 above observations suggest that, if we superpose 
 the low-lying pair vibration and the pair rotation in
 $^{134}$Sn, a resultant transition density may resemble 
 to that of the pair rotation in $^{142}$Sn. The analysis of
 other isotopes $^{136-140}$Sn, not shown here, also suggests the same
 feature.  We thus 
 deduce that the two pairing collectivities in $^{134-140}$Sn, i.e.
  the series i) and ii), merge into a single mode, which appears as the
 pair rotation in $^{142-150}$Sn.

\section{conclusions}

 We have investigated the monopole pair-additional transfer modes
 using the Skyrme-Hartree-Fock-Bogoliubov theory and the 
 linear response formalism of QRPA using the Skyrme parameter set SLy4
 and the DDDI pairing interaction. We particularly 
 analyzed the microscopic structure of the pairing
 collective modes in detail by evaluating 
 the forward and backward amplitudes $X_{ii'}^{\nu}$ and
 $Y_{ii'}^{\nu}$ of the QRPA phonon operator, and by 
 decomposing the transition densities associated with the
 pair-addition transfer operator.
 We apply this
 analysis for the low-lying pair vibration, the high-lying giant pair
 vibration and the pair rotation in even-even $A=120-150$ Sn isotopes.

 We have first investigated 
 the low-lying pair-addition vibration emerging in neutron-rich
 $^{134-140}$Sn in order to reveal its very large pair-addition
 strength\cite{Shimoyama11}. 
 The largest two-quasiparticle components of this mode are those
 involving the neutron quasiparticle states located just above the
 $N=82$ shell gap, i.e. 
 $2f_{7/2}, 3p_{3/2}, 3p_{1/2}, 2f_{5/2}, 1h_{9/2}$ etc, but it 
 turned out that contributions from these
 largest components can account for only a part of the amplitude and
 the long tail of the transition density.
 We found significant contributions from 
 the quasiparticle states which have larger orbital angular
 momenta $l > 5$, reaching to $l \gesim 10$, all of which
 are unbound continuum states. These two-quasiparticle configurations
 of high-$l$ states accumulate coherently to build up the large
 and extended tail of the transition density. 
 This suggests that a neutron pair transfered in the
 low-lying pair-addition vibration of $^{134-140}$Sn exhibits
 a strong spatial correlation, the di-neutron correlation,
 especially outside the nuclear surface.

 We have analyzed also isotopes $^{110-130}$Sn in order to
 explore the long-tail pairing vibration 
 in isotopes  closer to the stability line.
 The present calculation predicts presence of the giant pair vibrations,
 with two peaks GPV1 and GPV2, in $110 \lesim A \le 130$ 
 with excitation energy $\hbar\omega_\nu \approx 8-20$ MeV.
 The pair-addition strengths of GPV1 and GPV2 are
 comparable to (but slightly smaller than) that
 of the low-lying pair vibration in $^{134-140}$Sn.
 The detailed analysis of the phonon amplitudes and the
 transition density revealed that these modes have 
 similar microscopic structures as that of the low-lying pair vibration 
 in $^{134-140}$Sn, and the di-neutron character and the long tail 
 are also seen.
 The di-neutron character is also seen in the pair-addition vibrations
 in $^{132}$Sn and the pair rotation in $^{134-150}$Sn.

 From the systematic analysis 
 performed for all the isotopes from the stable ones to very neutron-rich 
 $^{150}$Sn, we found the above pairing collective modes are all related.
 The giant pair vibration GPV1 in $^{110-130}$Sn
 is smoothly connected to the lowest pair-addition vibration 
 in $^{132}$Sn (the lowest one populating the ground state of 
 $^{134}$Sn), which is then connected 
 to the pair rotation in $^{134-140}$Sn. A parallel series is
 the GPV2 in $^{120-130}$Sn, the second lowest pair-addition vibration 
 in $^{132}$Sn, and the low-lying pair vibration in $^{134-140}$Sn,
 which are also connected smoothly with changing $N$.
 At $^{142}$Sn, these two series 
 merge into a single mode, the pair rotation, which then continues further
 for $A>142$. The tail enhancement in the transition density and
 the collectivity in the pair-addition strength
 of these modes increase constantly with $N$, especially
 at $A>132$ where the neutron Fermi energy becomes very shallow.

\section*{ACKNOWLEDGMENTS}

 The authors thank K.~Yoshida for useful advises and discussions.
 The authors also thank K.~Matsuyanagi, K.~Yabana, T.~Nakatsukasa and
 H.~Liang for valuable discussions. 
 The numerical calculations were performed partly on the HITACHI SR1600
 supercomputer systems at Yukawa Institute for Theoretical Physics,
 Kyoto University.
 This work is supported by Grants-in-Aid for Scientific Research (No.
 21340073,
 No. 23540294, No.24105008, and No.25287065) from the Japan Society for
 the Promotion of Science.

\end{document}